\documentclass{elsart}
\usepackage{epsfig}
\usepackage{here}
\journal{Astroparticle Physics}

\def\nue{\nu_e}

\def\anue{\bar{\nu}_e}
\def\nux{\nu_x}
\newcommand{\be}{\begin{equation}}
\newcommand{\ee}{\end{equation}}
\newcommand{\eV}{{\rm eV}}

\usepackage{amssymb}

\begin{document}

\begin {frontmatter}

\title {Study of the effect of neutrino oscillations on the supernova neutrino signal in the LVD detector}

%%LVD Collaboration

\author[inr]{N.Yu.Agafonova},
\author[to]{M.Aglietta},
\author[bo]{P.Antonioli},
\author[bo]{G.Bari},
\author[inr]{V.V.Boyarkin},
\author[to]{G.Bruno},
\author[to]{W.Fulgione},
\author[to]{P.Galeotti},
\author[cf]{M.Garbini $^{\rm c,}$},
\author[lngs]{P.L.Ghia $^{\rm b,}$}, 
\author[bo]{P.Giusti},
\author[br]{E.Kemp},
\author[inr]{V.V.Kuznetsov},
\author[inr]{V.A.Kuznetsov},
\author[inr]{A.S.Malguin},
\author[bo]{H.Menghetti},
\author[bo]{A.Pesci},
\author[mit]{I.A.Pless},
\author[to]{A.Porta},
\author[inr]{V.G.Ryasny},
\author[inr]{O.G.Ryazhskaya},
\author[to]{O.Saavedra},
\author[bo]{G.Sartorelli},
\author[bo]{M.Selvi\corauthref{cor}}
\corauth[cor]{Corresponding author: Marco Selvi, c/o INFN - Sezione di Bologna, via Irnerio 46, 40126 Bologna - Italy, email: selvi@bo.infn.it},
\author[to]{C.Vigorito},
\author[lngs]{F.Vissani},
\author[lnf]{L.Votano},
\author[inr]{V.F.Yakushev},
\author[inr]{G.T.Zatsepin},
\author[bo]{A.Zichichi}

\address[inr]{Institute for Nuclear Research, Russian Academy of Sciences, Moscow, Russia}
\address[to]{Institute of Physics of Interplanetary Space, INAF, Torino, University of Torino and INFN-Torino, Italy}
\address[bo]{University of Bologna and INFN-Bologna, Italy}
\address[cf]{Museo Storico della Fisica, Centro Studi e Ricerche "E. Fermi", Rome, Italy} 
\address[lngs]{INFN-LNGS, Assergi, Italy}
\address[br]{University of Campinas, Campinas, Brazil}
\address[mit]{Massachusetts Institute of Technology, Cambridge, USA}
\address[lnf]{INFN-LNF, Frascati, Italy}

\begin{abstract}
The LVD detector, located in the INFN Gran Sasso National Laboratory (Italy), studies supernova neutrinos through the interactions with protons and carbon nuclei in the liquid scintillator and interactions with the iron nuclei of the support structure. 
We investigate the effect of neutrino oscillations in the signal expected in the LVD detector. The MSW effect has been studied in detail for neutrinos travelling through the collapsing star and the Earth.
We show that the expected number of events and their energy spectrum are sensitive to the oscillation parameters, in particular to the mass hierarchy and the value of $\theta_{13}$, presently unknown. 
Finally we discuss the astrophysical uncertainties, showing their importance and comparing it with the effect of neutrino oscillations on the expected signal.
%Finally we discuss the impact of the astrophysical uncertainties on the expected signal.
\end{abstract}


\end{frontmatter}

\normalsize 

\section{Introduction}
%Neutrino conversion among flavors has been discovered and firmly established in the recent few years studying atmospheric \cite{skatm}, solar \cite{Cl} \cite{Ga} \cite{Sage} \cite{Kam} \cite{sksol} \cite{sno}, reactor \cite{KamLAND} and accelerator \cite{K2K} neutrinos.
There are many experimental works suggesting neutrino conversion
among flavors in the recent few years, through the study of atmospheric \cite{skatm}, solar \cite{Cl} \cite{Ga} \cite{Sage} \cite{Kam} \cite{sksol} \cite{sno}, reactor \cite{KamLAND} and accelerator \cite{K2K} neutrinos.
The interpretation of all these phenomena in terms of 
neutrino oscillations is rather robust, because it is able to include all the experimental data (except the ``not yet confirmed'' LSND \cite{lsnd} signal), even if the expected oscillatory behavior, in terms of the observable $L/E$, has not been yet experimentally observed (preliminary results that show a low significance hint for a oscillatory behavior have been found by a re--analysis of the SK data \cite{skatm}). 
An interesting fact is that the inclusion of the MSW effect \cite{msw} 
permits a consistent interpretation of KamLAND results and the `high 
energy' solar neutrino data \cite{Kam,sksol,sno}.

In the standard three flavor scenario, six parameters must be determined by oscillation experiments: 3 mixing angles ($\theta_{{\rm sol}}$, $\theta_{13}$, $\theta_{{\rm atm}}$ ), 2 squared mass differences ($\Delta m^2_{{\rm sol}}$ and $\Delta m^2_{{\rm atm}}$) and 1 CP-violation phase $\delta$. 
A recent analysis of all the available experimental data \cite{VissStrum05}  constrains the ``atmospheric'' and ``solar'' parameters to be in the following $99\% ~C.L.$ ranges (compare also with the results in \cite{FLglobal}): 

\begin{table}[h]
$$\begin{array}{lrlc}
\hbox{Oscillation parameter}&\multicolumn{2}{c}{\hbox{central value}} &\hbox{$99\%$ C.L. range}\\  \hline
\hbox{solar mass splitting} & \Delta m^2_{{\rm sol}} ~= & (8.0\pm 0.3)\,10^{-5}~\eV^2 &  (7.2\div 8.9)\,10^{-5}\eV^2 \\
\hbox{atm. mass splitting~~~~}  & |\Delta m^2_{{\rm atm}}| ~= & (2.5\pm 0.3) \,10^{-3}~\eV^2~~~ & (1.7\div 3.3) \,10^{-3}\eV^2\\ 
\hbox{solar mixing angle} &  \tan^2 \theta_{{\rm sol}} ~=&  0.45\pm0.05 & 30^\circ < \theta_{{\rm sol}}<38^\circ \\
\hbox{atm. mixing angle} & \sin^2 2\theta_{{\rm atm}} ~= &   1.02\pm 0.04 &36^\circ <\theta_{{\rm atm}}< 54^\circ\\
\end{array}$$
%\caption{\em Summary of present information from oscillation data. A $99\%$ C.L.\ range is a $2.58\sigma$ range.
\label{tab1}
\end{table}

However the other parameters are not completely determined: the $\theta_{13}$ mixing angle is only upper limited, mainly by the Chooz experiment data \cite{Chooz} ($\sin^2 \theta_{13} < 3.~ 10^{-2}$  at the $99 \%~ C.L.$), the sign of $\Delta m^2_{{\rm atm}}$ (that fixes the so--called mass hierarchy) is completely unknown, as well as the CP--violation phase $\delta$.

Because of the wide range of matter density in the stellar envelope,
a supernova explosion represents a unique scenario for further study of the neutrino oscillation mixing matrix. Indeed neutrinos can cross two resonance density layers and therefore the resulting possible mixing scenarios are different from the solar ones. The emerging neutrino spectra are sensitive to the sign of $\Delta m^2_{{\rm atm}}$ and to the value of $\theta_{13}$.

Before proceeding, it is important to recall that, at present, there is
not a unique theory of supernova explosions. Till now, numerical
investigations of the ``standard model'' based on a delayed scenario of the
explosion 
%Bethe-Wilson
\cite{Olga0} failed to reproduce the explosion. On top of
that, other models are being studied where rotation \cite{Olga1} or
magnetic field \cite{Olga2} play an essential role. In the
following, we will use a simple description of the neutrino flux that does
not contradict the SN1987A events seen by Kamiokande-II \cite{Olga3}, IMB \cite{Olga4} and Baksan \cite{Olga5}, see e.g. \cite{Olga6} for a discussion, although it is not able to
take into account the events seen in Mont-Blanc observatory \cite{Olga7,Oscar0}. This ``standard''
description, however, corresponds to the expected neutrino emission in the
delayed scenario and in the last phase of the collapse with rotation \cite{Olga8}. For this reason, we take it as a useful starting
point for the investigation of the impact of oscillations in the neutrino signal
from a supernova.

The main aim of this paper is to show how neutrino oscillations 
affect the signal detected by the LVD observatory in the INFN Gran Sasso National Laboratory, Italy.
We also evaluate the impact on the signal of the astrophysical parameters of the supernova explosion mechanism, such as the total energy emitted in neutrinos, the star distance, the neutrino--sphere temperatures and the partition of the energy among the neutrino flavors.

In section \ref{se:sn} we describe the characteristics of the neutrino fluxes emitted during a gravitational core collapse. In section \ref{se:osc} the neutrino oscillation mechanism is shown, in particular the peculiarities of the MSW effect in the supernova matter and in the Earth. The LVD detector and the relevant neutrino interactions both in the liquid scintillator and in the iron support structure are described in section \ref{se:lvd}. The impact of neutrino oscillations in the signal expected in the LVD detector is presented in section \ref{se:res} while the uncertainties in the astrophysical parameters and their effect on the results are discussed in section \ref{se:ap}. Finally, the conclusions are drawn in section \ref{se:sum}.
Two appendices complete this work, describing in more detail the MSW effect calculation in the Earth (A) and the neutrino interaction with the iron of the LVD support structure (B).

Preliminary results have been presented previously in \cite{taup01}, \cite{icrc03} and \cite{Giacobbe}.

\section{Supernova neutrino emission}
\label{se:sn}
At the end of its burning phase a massive star ($M \ge 8 M_\odot$) explodes into a supernova, originating a neutron star which rapidly cools down (order of tens of seconds) emitting about 99\% of the liberated gravitational binding energy in neutrinos.

The time--integrated spectra can be well approximated by the
pinched Fermi--Dirac distribution.
%, with an effective degeneracy parameter $\eta$.
For the neutrinos of flavor $\alpha$, we have
\begin{eqnarray}
F^0_\alpha(E,T_\alpha,\eta_\alpha,L_\alpha,D) =
\frac{L_\alpha}{4\pi D^2 T^4_\alpha F_3(\eta_\alpha)} \frac{E^2}{e^{E/T_{\alpha}-\eta_\alpha}+1} 
\label{eq:spectra}
\end{eqnarray}
where $D$ is the distance to the supernova, $E$ is the neutrino energy, $L_\alpha$ is the time--integrated
energy of the flavor $\nu_\alpha$, $T_\alpha$ represents the
effective temperature of the $\nu_\alpha$ gas inside the star, $\eta_\alpha$ is the pinching parameter, $F_3(\eta_\alpha) \equiv \int_0^\infty x^3 / (e^{x-\eta_\alpha}+1) ~dx$ is the normalization factor.
In most of this work we assume for simplicity that $\eta=0$ for all 
neutrino flavors; this choice results in the relation $\langle E_\alpha \rangle \simeq 3.15 ~T_\alpha$ between the mean neutrino energy temperature.

Due to different trapping processes, the neutrino
flavors originate in layers of the supernova with different
temperatures. The electron (anti)neutrino flavor is kept in thermal
equilibrium by $\beta$ processes up to a certain radius usually referred to as
the ``neutrino--sphere'', beyond which the neutrinos stream off
freely. However, the practical absence of muons and taus in the
supernova core implies that the other two neutrino flavors, here
collectively denoted by $\nu_x$ ($\nu_\mu, \nu_\tau, \bar\nu_\mu,
\bar\nu_\tau$), interact primarily by less efficient neutral--current processes.
Therefore, their spectra are determined at deeper, $i.e.$ hotter,
regions. In addition, since the content of neutrons is larger
than that of protons, $\nu_e$'s escape from more external regions than
$\bar{\nu}_e$'s. 
This rough picture leads to the hierarchy
$\langle E_{\nu_e}\rangle < \langle E_{\bar\nu_e}\rangle < \langle
 E_{\nu_x}\rangle $.
Typical ranges for 
the average energies of the time--integrated neutrino
spectra obtained in simulations
are $\langle E_{{\nu}_e} \rangle = 10-12$ MeV, $\langle E_{\bar{\nu}_e} \rangle
= 11-17$ MeV, and $\langle E_{{\nu}_x} \rangle =15-24$ MeV
\cite{Janka:1992jk,Totani:1998vj}.
However, 
recent studies with an improved treatment of $\nu$ transport,
micro--physics, the inclusion of the nucleon bremsstrahlung, and the
energy transfer by recoils, find somewhat smaller differences between the
$\bar\nu_e$ and $\nu_x$ spectra \cite{Janka}.

The amount of the total binding energy $E_b$ taken by each flavor is 
$L_\alpha = f_{\nu_\alpha}~ E_b$, with $f_{\nu_e}=17-22\%, 
f_{\bar\nu_e}=17-28\%, f_{\nu_x}=16-12\%$ (see e.g. \cite{VissProbes} ). 
Thus, the so--called ``energy equipartition'' has to be intended as ``within a factor of two'' \cite{Janka}.

\begin{figure}[t]
    \begin{center}
      \includegraphics[height=36pc]{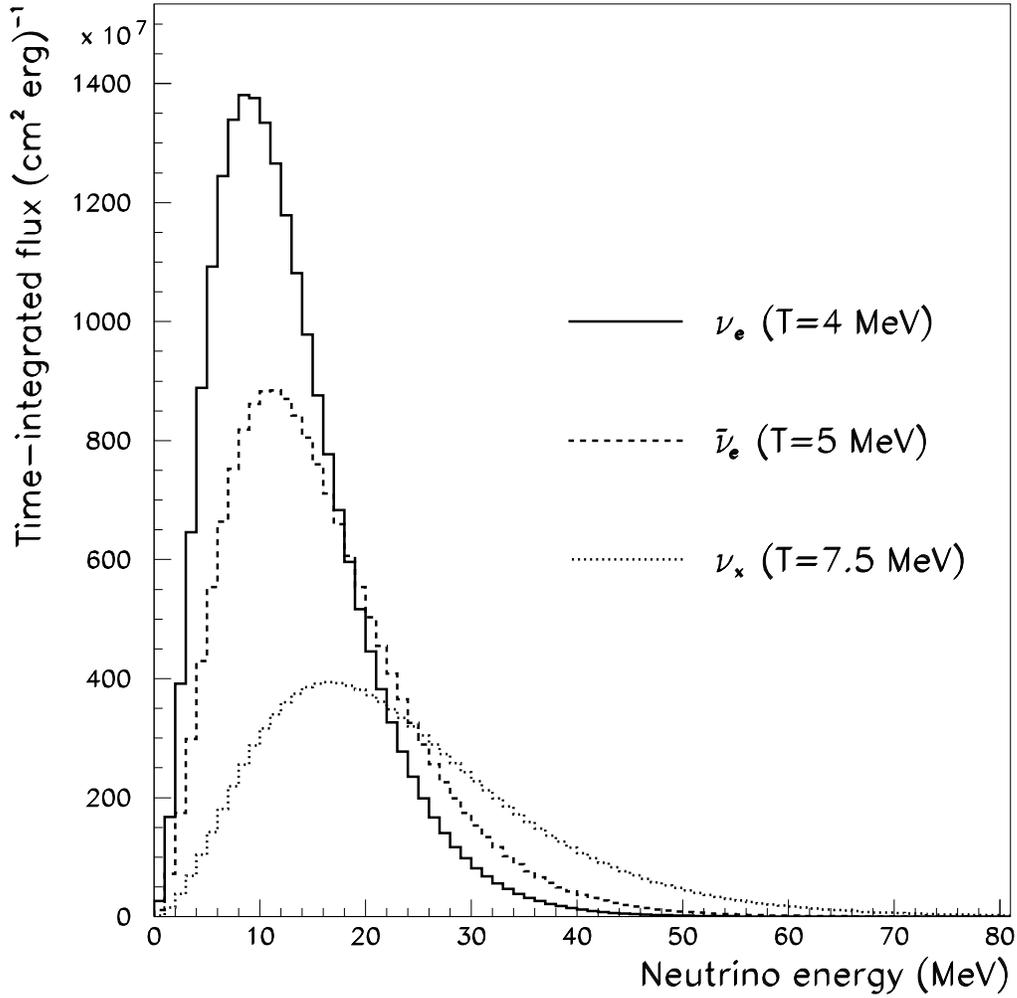}
    \end{center}
    \vspace{-.5cm}
    \caption{Neutrino energy spectra at the neutrino--sphere.}
    \label{fi:flux}
    \vspace{.5cm}
\end{figure}

In the following, if not specified differently, we assume a galactic supernova explosion at a typical
distance of $D = 10$~kpc, with a total binding energy of $E_b = 3 \cdot
10^{53}$ erg and perfect energy equipartition $f_{\nu_e}=f_{\bar\nu_e}=f_{\nu_x}=1/6$.
We also assume that the fluxes of
$\nu_\mu$, $\nu_\tau$, $\bar\nu_\mu$, and $\bar\nu_\tau$ are identical; we
fix 
$T_{\nu_x} / T_{\bar{\nu}_e} = 1.5$, 
$T_{\nu_e} / T_{\bar{\nu}_e}  = 0.8$
and $T_{\bar{\nu}_e} 
%\in \{4,~7\}~ 
=5~{\rm MeV}$ \cite{Janka}.
With these assumptions the resulting neutrino energy spectra generated inside the star are shown in figure \ref{fi:flux}.

We postpone to section \ref{se:ap} a discussion about the implications to the expected number of events in the LVD detector due to the uncertainties in the astrophysical parameters. 

\section{Neutrino flavor transition in the star and in the Earth.}
\label{se:osc}
In the study of supernova neutrinos,
$\nu_{\mu}$ and $\nu_{\tau}$ are indistinguishable, both in the star and in
the detector, because of the corresponding charged lepton production threshold; consequently, in the frame of three--flavor oscillations,
the relevant parameters are just
$(\Delta m^2_{{\rm sol}}, U_{e2}^2)$ and 
$(\Delta m^{2}_{\rm atm}, U_{e3}^2)$ \footnote{$U_{e1}^2=\cos^2 \theta_{13} \cdot \cos^2 \theta_{12} \simeq \cos^2 \theta_{12}$,  $U_{e2}^2=\cos^2 \theta_{13} \cdot \sin^2 \theta_{12} \simeq \sin^2 \theta_{12}$ and $U_{e3}^2 = \sin ^2 \theta_{13}$.}.

We will adopt the following numerical values:
$\Delta m^2_{{\rm sol}}=8 \cdot 10^{-5}~{\rm eV}^2$, 
$\Delta m^{2}_{\rm atm}=2.5 \cdot 10^{-3}~ {\rm eV}^2$, 
$U_{e2}^2=0.33;$ the selected solar parameters 
$(\Delta m^2_{{\rm sol}}, U_{e2}^2)$ describe the
LMA solution, as it results from a global analysis including solar, CHOOZ and KamLAND $\nu$ data \cite{VissStrum05}.

\begin{figure}[t]
    \begin{center}
      \includegraphics[width=14cm]{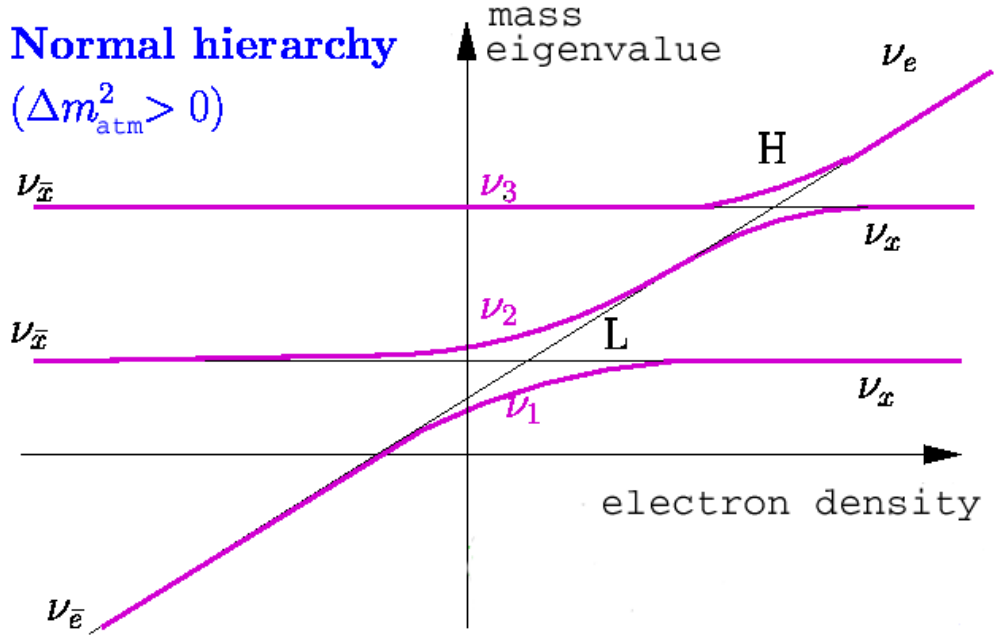}
      \includegraphics[width=14cm]{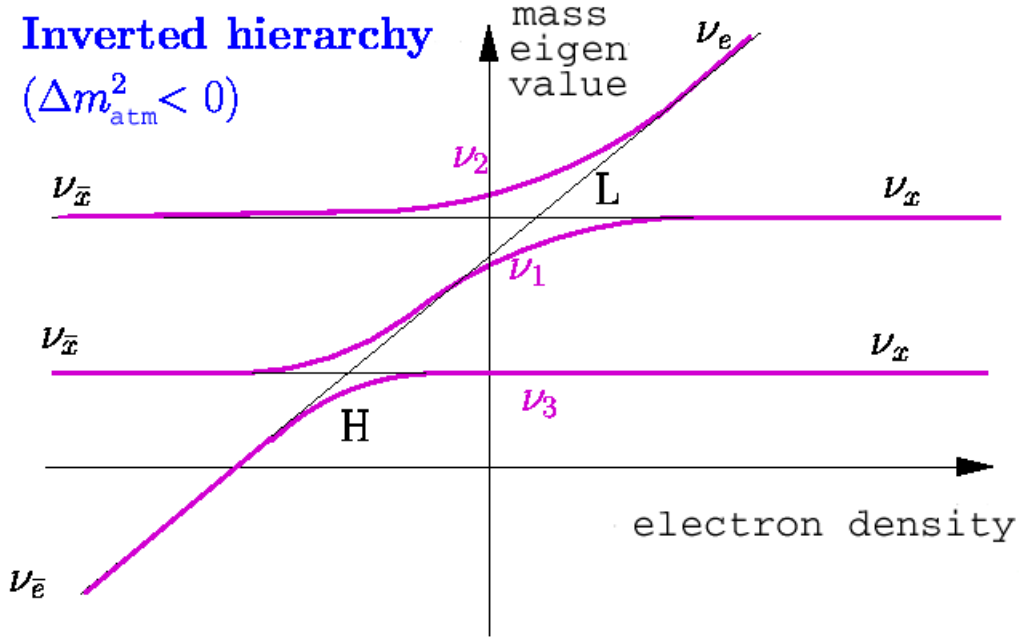}
    \end{center}
%    \vspace{-1.cm}
    \caption{Crossing level scheme for Normal ({\it top}) and Inverted ({\it bottom}) hierarchy. Solid thick purple lines show the eigenvalues of the effective Hamiltonian as function of the electron number density. The thin black lines correspond to the energy of the flavor levels $\nue$ and $\nux$. Negative values of the electron number density are related to the antineutrino channel.}
    \label{fi:crossNH}
\end{figure}

As described in figure \ref{fi:crossNH} neutrinos, in the normal mass hierarchy (NH) scheme, cross two so--called Mikheyev--Smirnov--Wolfenstein \cite{msw} resonance layers
in their path from the high density region where they are generated to the lower density one where they escape the star:
one at higher density (H), which corresponds to $(\Delta m^{2}_{\rm atm}, U_{e3}^2)$ and $\rho=300 \div 6000~\mbox{g/cm}^3$ \footnote{the values are respectively for $E_\nu$ equal to $100$ and $5$ MeV}, and
the other at lower density (L), corresponding to 
$(\Delta m^{2}_{{\rm sol}}, U_{ e2}^2)$ and $\rho=5 \div 100~\mbox{g/cm}^3$. Antineutrinos do not cross any MSW resonance \cite{Panta,Dutta,Dighe}.

For inverted mass hierarchy (IH), transitions at
the higher density layer occur in the $\bar \nu$
sector, while at the lower density layer they occur 
in the $\nu$ sector. 

Neutrinos are originated in regions of the star where the density is very high, so that the effective mixing matrix in matter is practically diagonal. Thus the created neutrino flavor eigenstate is completely projected into one neutrino mass eigenstate (represented by the thick purple line in figure \ref{fi:crossNH}). Then the neutrino starts its path through the matter to escape the star. If the matter density changes in a smooth way, then the propagation is said to be ``adiabatic''. It means that the neutrino propagates through the star being the same mass eigenstate (i.e., referring to figure \ref{fi:crossNH}, staying over the same thick purple line). The adiabaticity condition depends both on the density variation and on the value of the oscillation parameters involved.

Given the energy range of supernova $\nu$ (up to $\sim 100~{\rm MeV}$)
and considering a star density profile $\rho \propto 1/r^3$, 
the L transition is adiabatic for any LMA solution values. Thus the probability to jump onto an adjacent mass eigenstate (hereafter called {\it flip} probability) is null  ($P_L=0$). The adiabaticity at the $H$ resonance
depends on the value of $U_{e3}^2$ in the following way \cite{Dighe}:
$$P_{\rm H} \propto \exp~[-~ const~ U_{e3}^2~ (\Delta m^{2}_{\rm atm} / E)^{2/3}~ ]$$ where $P_{\rm H}$ is the flip probability at the H resonance.
 
When $U_{e3}^2 \geq 5 \cdot 10^{-4}$ 
the conversion is completely adiabatic ({\it ad}) and the flip probability is null ($P_{\rm H}=0$); conversely, when $U_{e3}^2 \le 5 \cdot 10^{-6}$ the conversion is completely non adiabatic ({\it na}) and the flip probability is $P_{\rm H}=1$. We used in the calculation 
$U_{e3}^2=10^{-2}$, which is just behind the corner of the CHOOZ upper limit, for the adiabatic case and $U_{e3}^2=10^{-6}$ for the non adiabatic one.

For neutrinos, in the NH-{\it ad} case $\nue$ generated in the star arrive at Earth as $\nu_3$, so their probability to be detected as $\nue$ is $U_{e3}^2 \sim 0$. Thus, the detected $\nue$ come from higher--energy $\nux$ in the star that get the Earth as $\nu_2$ and $\nu_1$. \\If the H transition is not adiabatic or if the hierarchy is inverted the original $\nue$ get the Earth as $\nu_2$ and their probability to be detected as $\nue$ is $U_{e2}^2 \sim 0.3$.

For antineutrinos, in the NH case or in the IH-{\it na}, the $\bar \nu_e$ produced in the supernova core arrive at Earth as $\nu_1$, and they have a high ($U_{e1}^2 \simeq 0.7$) probability to be detected as $\bar\nu_e$. 
On the other hand, the original $\bar \nu_x$ arrive at Earth as $\nu_2$ and $\nu_3$ and are detected as $\bar \nu_e$ with probability $U_{e2}^2$.\\ 
In the IH-{\it ad} case the detected $\bar \nu_e$ completely come from the original, higher--energy $\bar \nu_x$ flux in the star.

The oscillations scheme can be summarized as:
\begin{eqnarray}
 F_e = & P_{\rm H} U_{e2}^2 F_e^0 & + ~(1-P_{\rm H} U_{e2}^2) F_x^0  
\label{eq:fenh} \\
F_{\bar e} = & U_{e1}^2 F_{\bar e}^0 & + ~ U_{e2}^2 F_{\bar x}^0  
%~~~~~~~~~~~~~ {\rm for~ normal~ hierarchy~ and }
\label{eq:faenh}
\end{eqnarray}
for normal hierarchy and 
\begin{eqnarray}
F_e = & U_{e2}^2 F_e^0 & + ~ U_{e1}^2 F_x^0  \label{eq:feih}\\
F_{\bar e} = & P_{\rm H} U_{e1}^2 F_{\bar e}^0 & + ~ (1 - P_{\rm H} U_{e1}^2) F_{\bar x}^0 
\label{eq:faeih}
\end{eqnarray}
 for inverted hierarchy, \\
where $F_{any}^0$ are the original neutrino fluxes in the star and $F_{any}$ are the observed $\nu$ fluxes. 
One can notice that, 
if the H transition is completely non adiabatic ($P_{\rm H}=1$), the NH and IH cases coincide.
Thus, to see any effect due to the mass hierarchy, the H transition must be adiabatic, i.e. $\theta_{13}$ has not to be too small. 

When we consider the effect of the Earth in the neutrino path to the detector, we must replace, in the detected flux estimation in formulas (\ref{eq:fenh}-\ref{eq:faenh}-\ref{eq:feih}-\ref{eq:faeih}), $U_{ei}^2$ with $P_{ie}~ (i=1,2)$, the probability for the mass eigenstate $\nu_i$ to be detected as $\nu_e$ (or $\anue$) after travelling through the Earth \cite{Lunard}, which depends on the solar oscillation parameters and on the travelled density profile through the Earth.
We developed a complete 3--flavor calculation, describing the Earth interior as made of 12 equal density steps, following the {\it Preliminary Reference Earth Model} matter density profile \cite{prem}. For each constant density step we compute the exact propagator of the evolution matrix and we get the global amplitude matrix by multiplying the propagators of the traversed density layers, as described e. g. in \cite{Akhmedov}.

More detail about the calculation of the probabilities $P_{ie}$ are given in Appendix \ref{AppA}.

\section{LVD detector and the observable neutrino interactions}
\label{se:lvd}

\begin{figure}[t]
  \begin{minipage}{.40\columnwidth}
%    \begin{center}
      \includegraphics[height=20pc]{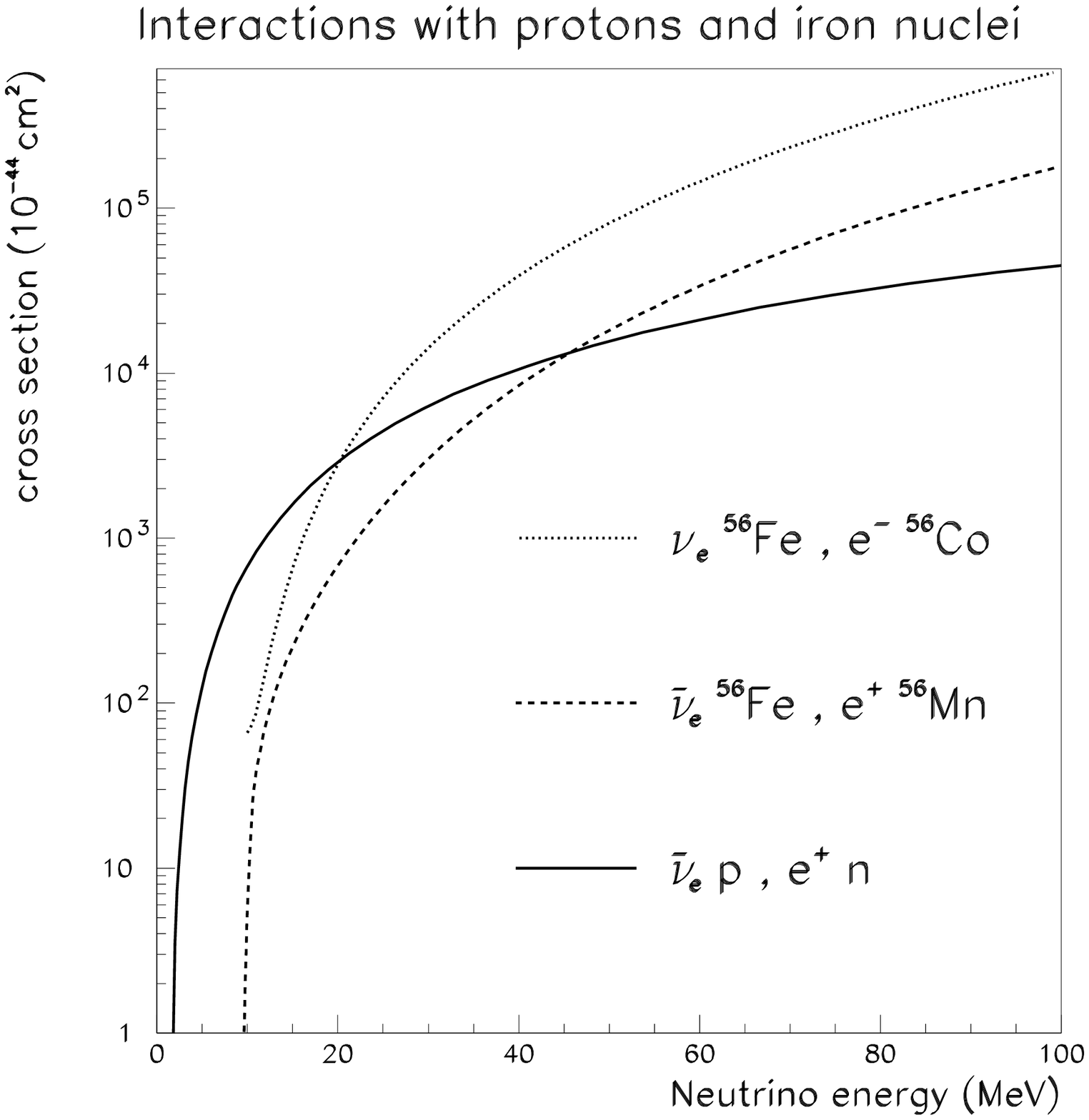}
%    \end{center}
%    \vspace{-1.cm}
  \end{minipage}
  \hspace{3.5pc} %%%%% space between two figures
  \begin{minipage}{.40\columnwidth}
      \includegraphics[height=20pc]{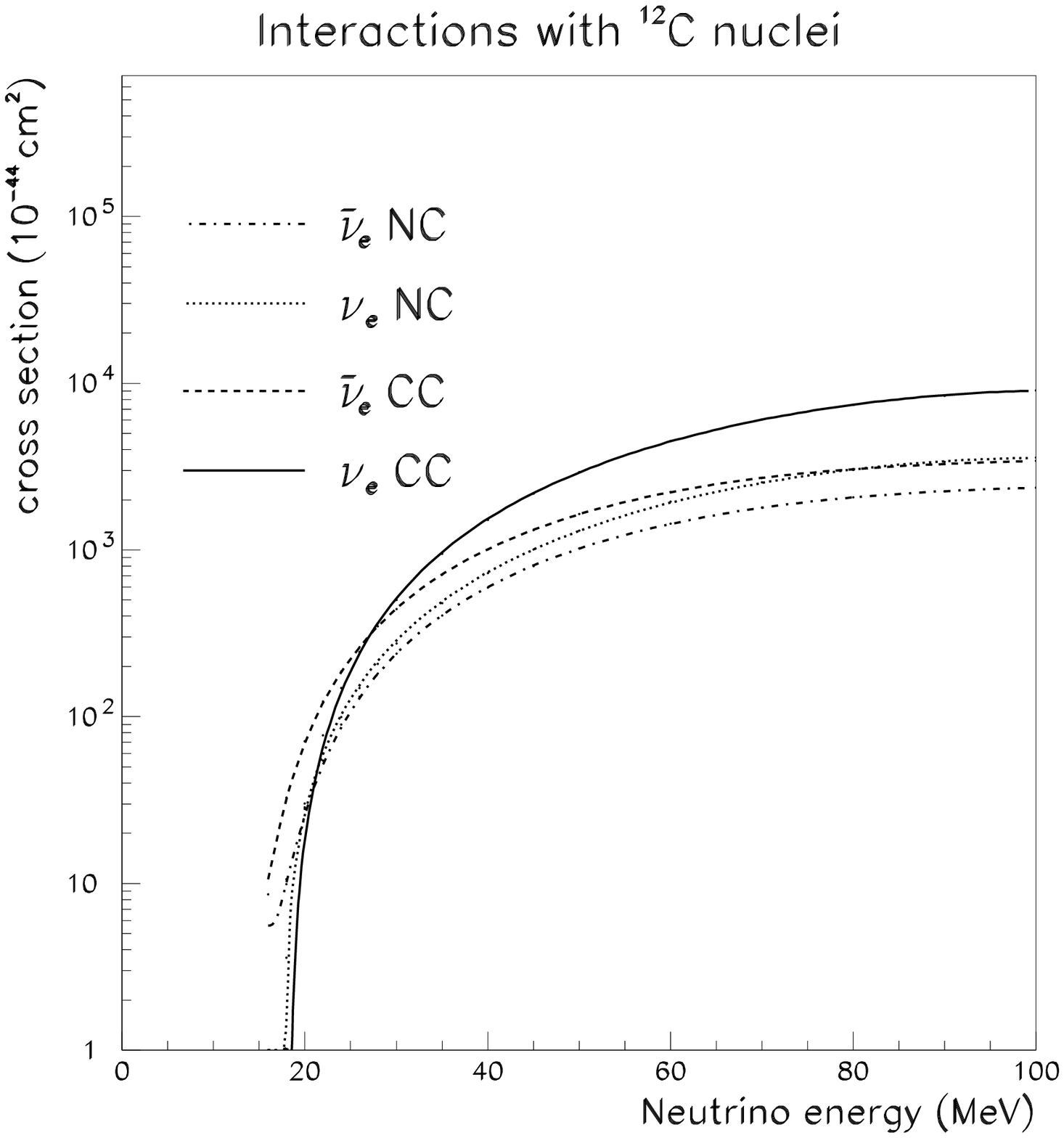}
  \end{minipage}
    \caption{Theoretical cross sections of the neutrino CC interactions with protons and iron nuclei ({\it left}) and CC and NC interactions with carbon nuclei ({\it right}).}
    \label{fi:sig}
\end{figure}

The Large Volume Detector (LVD) in the INFN Gran Sasso National Laboratory,
Italy, consists of an array of 840  
liquid scintillator counters, 1.5 m$^3$ each,
arranged in a compact and modular geometry; a detailed description can be found in \cite{LVD,StatRep}.
The active scintillator mass is $M=1000$ t. 
The counter are operated at energy threshold ${\mathcal E}_h\simeq 5~ {\rm MeV}$.
To tag the delayed $\gamma$ pulse due to the
neutron capture, all counters are equipped with an additional discrimination
channel, set at a lower
threshold, ${\mathcal E}_l\simeq 1~ {\rm MeV}$.
The energy resolution is $\sigma_{E}/{E} = 0.07 + 0.23\cdot ({E} /\rm{MeV})^{-0.5}$

\vspace{1. cm}

The observable neutrino reactions that occur in the scintillator are:
\begin{enumerate}
\item
$\bar\nu_e~ p,e^+ n$, (physical threshold $E_{\bar\nu_e} > 1.8$ MeV) observed through a prompt signal from ${e}^+$ above
threshold ${\mathcal E}_h$ (detectable energy $E_d \simeq E_{\bar\nu_{e}}-1.8$ {\rm MeV}
$+ 2~ m_e c^2 $), followed by the signal from the ${n ~p,d}~ \gamma$ capture
($E_{\gamma} = 2.2$ {\rm MeV}) above ${\mathcal E}_l$ and with a mean delay
$\Delta t \simeq 185~\mu \mathrm{s}$ (as it comes out from estimation \cite{Olga9} and Montecarlo simulation \cite{AmandaPhD}).  The cross section for this reaction has been recalculated \cite{VisStrum} with a better treatment of the $10-100~{\rm MeV}$ region, i.e. the supernova neutrino energy.
The cross section behavior with energy is shown in figure \ref{fi:sig} ({\it left} plot, solid line). 
The total number of free protons in the scintillator is $9.34~\cdot 10^{31}$.

\item 
\noindent $\nu_e\,^{12}\mathrm{C},^{12}\!\mathrm{N}\ e^-$, (physical threshold $E_{\nu_e} > 17.3$ MeV)
observed through two signals: the
prompt one due to the $e^-$ above ${\mathcal E}_h$ (detectable energy
$E_d \simeq E_{\nu_e}-17.3$ MeV) followed by the signal, above ${\mathcal E}_h$, from the
$\beta^+$ decay of $^{12}\mathrm{N}$ (mean life time $\tau = 15.9$ ms).
The efficiency for the detection of the $^{12}\mathrm{N}$ beta decay product is $90\%$ \cite{pietroCC,TesiPorta}.

\item 
\noindent  $\bar\nu_e\, ^{12}\rm{C},^{12}\!\rm{B}\ e^+$, (physical threshold $E_{\bar\nu_e} > 14.4$ MeV) observed
through two signals: the prompt one due to the $e^+$ (detectable energy
$E_d \simeq E_{\bar\nu_e}-14.4\, \mathrm{MeV} + 2~m_e c^2$) followed by the signal
from the $\beta^-$ decay of $^{12}{\rm B}$ (mean life time $\tau= 29.4$ ms).
As for reaction {\em (2)}, the second signal is detected above the threshold
${\mathcal E}_h$ and the detection efficiency of the $^{12}\mathrm{B}$ beta decay product is $75\%$ \cite{pietroCC,TesiPorta}.

\item 
\noindent  $\stackrel{\scriptscriptstyle (-)}{\nu}_{\!\alpha}\,$
${}^{12}{\rm C},\stackrel{\scriptscriptstyle (-)}{\nu}_{\!\alpha}\,$
${}^{12}{\rm C}^*$
($\alpha=e,\mu,\tau$), (physical threshold $E_\nu > 15.1$ MeV) whose signature is the monochromatic photon from
carbon de--excitation ($E_{\gamma}=15.1$ MeV), above ${\mathcal E}_h$, detected with $55\%$ efficiency \cite{ANT91}. \\
Cross sections for reactions $(2),~(3)$ and $(4)$ are taken from \cite{CrossC} and shown in figure \ref{fi:sig} ({\it right}).

\item 
\noindent $\stackrel{\scriptscriptstyle (-)}{\nu}_{\!\alpha}\,$
$e^-,\stackrel{\scriptscriptstyle (-)}{\nu}_{\!\alpha}\, e^-$, which yields a single
signal, above ${\mathcal E}_h$, due to the recoil electron. Because of the low number of expected events (about a dozen) and the lack of a clear pattern for this interaction, we will not consider it in the following.
\end{enumerate}

The iron content in LVD (about 900 t) is concentrated in two components:
the stainless steel tank (mean thickness: $0.4$ cm) which contains 
the liquid scintillator and the iron module (mean thickness: $1.5$ cm) which hosts a cluster 
of 8 tanks. Indeed, the higher energy part of the $\nu$ flux can be detected 
also with the $\nu~ (\bar\nu)  
~{\rm Fe}$ interaction, which results in an 
electron (positron) that can exit iron and release energy in the scintillator. 
The considered  reactions are:
\begin{enumerate}
\setcounter{enumi}{5}
\item \noindent  $\nu_e\,^{56}\mathrm{Fe},^{56}\!\mathrm{Co}^*\ e^-$. The mass difference between the nuclei is $\Delta_{m_n}=m_n^{\rm{Co}} - m_n^{\rm{Fe}} = 4.055~{\rm MeV}$ and the first $\rm{Co}$ allowed state is at $1.72~{\rm MeV}$.
Other allowed levels are present in Cobalt, as shown in fig. \ref{fi:Co}, whose energy $E_{\rm level}$ is $3.59$, $7.2$, $8.2$, $10.59$ MeV. 
 Indeed, 
the electron kinetic energy is  $E_{e^-} = (E_{\nu_e} - \Delta_{m_n} - E_{\rm level} - m_e)$ .
Moreover, some gamma rays are produced in the interaction, depending on the excitation level considered.

A full simulation of the LVD support structure and of the scintillator detectors has been developed in order to get the efficiency for electron and gammas, generated randomly in the iron structure, to reach the scintillator with energy higher than ${\mathcal E}_h$. It is greater than $20\%$ for $E_\nu > 30~{\rm MeV}$ and grows up to $70\%$ for $E_\nu > 100~{\rm MeV}$.
On average, the electron energy detectable is $E_d \simeq 0.40 \times E_\nu$. The total number of iron nuclei is $9.22~\cdot10^{30}$.

\item \noindent  $\bar\nu_e\,^{56}\mathrm{Fe},^{56}\!\mathrm{Mn}\ e^+$,
the energy threshold is very similar to that of reaction {\em (6)}. In this work, for simplicity, the same efficiency is assumed.
\end{enumerate}

The total cross section for reactions {\em (6),(7)} are taken respectively from \cite{Kolbe1} and \cite{Kolbe2} and plotted in figure \ref{fi:sig} ({\it left}), while the probability to select a particular Cobalt excitation level is taken from \cite{Kurta}.
More detail about the neutrino--iron cross section, the Cobalt energy levels and the simulation of the interactions in the LVD detector are described in Appendix \ref{AppB}.

It is necessary to point out that, up to now, we calculated
only $\nu$-Fe charged current interactions. 
The estimation of $\nu$-Fe neutral current interaction cross section shows that they are roughly $30 \%$ of the CC ones \cite{Kurta}. 
They should be taken into account in future works.

The number of all the possible targets present in the LVD detector is listed in table \ref{ta:targ}.

\begin{table}[h]
 \caption{Number of targets in the LVD detector. \label{ta:targ}}
     \vspace{.3 cm}
\begin{center}
\begin{tabular}{|l|c|c|c|}
\hline
Target Type & Contained in &       Mass (t)       & Number of targets      \\
\hline
Free protons         & Liquid Scintillator     & $1000$ & $9.34~10^{31}$   \\
Electrons     & `` & $1000$     & $3.47~10^{32}$  \\
C Nuclei     & `` &    $1000$  & $4.23~10^{31}$      \\
Fe Nuclei     & Support Structure & $900$     & $9.71~10^{30}$   \\
\hline
\end{tabular}
\end{center}
\end{table}

\section{Expected neutrino signals}
\label{se:res}
The number of events detected during the supernova explosion is calculated as:

\begin{equation}
N_{ev} = N_t \cdot \int_{0}^{\infty} F(E_\nu) \cdot \sigma(E_\nu) \cdot \epsilon(E_\nu) ~ dE_\nu
\end{equation}

where 
$N_t$ is the number of target nuclei, $F$ is the neutrino flux, $\sigma$ is the interaction cross section, $E_\nu$ is the neutrino energy and $\epsilon$ is the efficiency for the detection of the interaction products, where the effect of the detector energy threshold is included.

In the following we show the effect of neutrino oscillations in the SN matter in the various interaction channels and the possible interplay among them. Then, the effect of the Earth matter is taken into account in the last subsection, considering only the inverse beta decay channel.

\subsection{Inverse beta decay}
\begin{figure}[b!] %checked 310106
    \begin{center}
      \includegraphics[height=36pc]{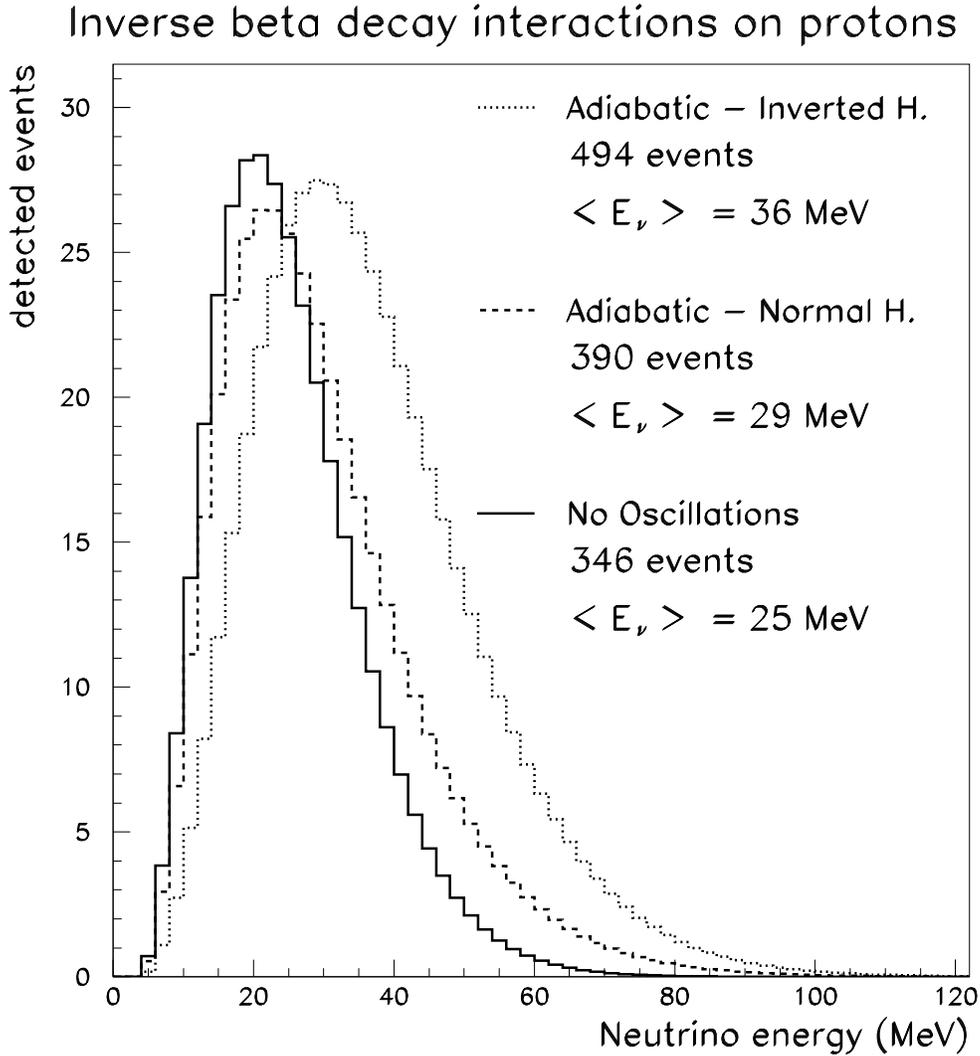}
    \end{center}
%    \vspace{-1.cm}
    \caption{Neutrino energy distribution in the $\bar\nu_e$ interactions with $p$ expected in LVD for three oscillation scenarios: no oscillation (solid line), adiabatic transition with NH (dashed), adiabatic transition with IH (dotted). The situation in the non adiabatic transition cases is identical to the adiabatic transition with NH case. The integral number of detected events is shown.}
    \label{fi:figibdosc}
%  \end{minipage}
\end{figure}

The main interaction in LVD is the inverse beta decay (IBD) of electron antineutrinos. In figure \ref{fi:figibdosc} we show the energy spectra of the detected neutrinos in the case of no oscillation and in the case of adiabatic transition with NH and IH. We remind here that the non--adiabatic transition case (for both NH and IH) is coincident with the adiabatic NH case.

\begin{figure}[h!]  %checked 310106
    \begin{center}
      \includegraphics[height=36pc]{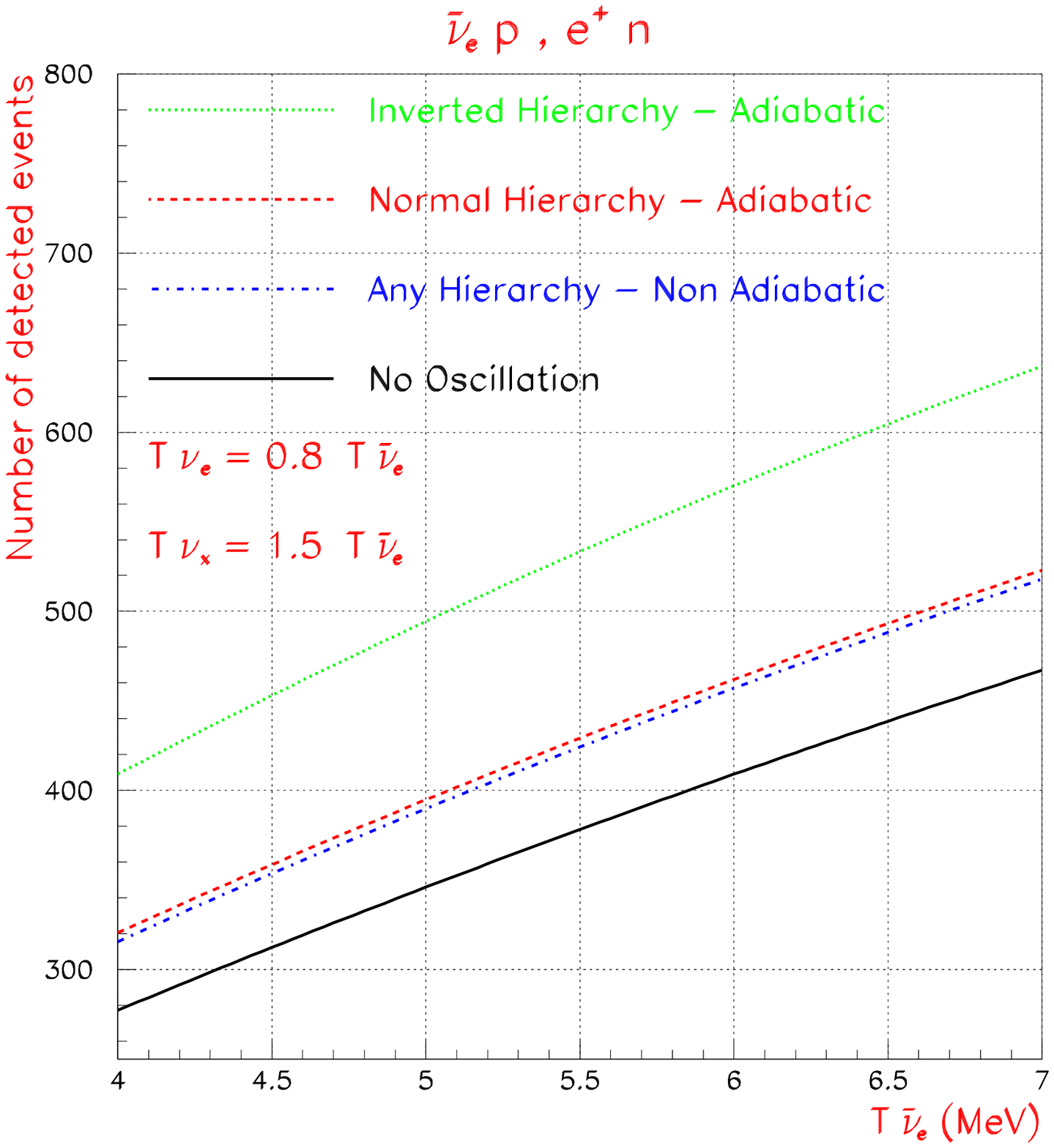}
    \end{center}
%    \vspace{-1.cm}
    \caption{Number of detectable $\bar\nu_e$ interactions with $p$ expected in LVD as a function of the $\bar\nu_e$ neutrino--sphere temperature.}
    \label{fi:repibd}
%  \end{minipage}
\end{figure}

In the case of oscillation, adiabatic, normal hierarchy, there is a contribution ($\sin^2\theta_{12}$) of the original higher--energy $\bar \nu_x$ which gives rise to a higher average neutrino energy and, due to the cross section increase, to a larger number of detected events.
The $\nu_x$ contribution is even higher ($\sim 1$) if the transition is adiabatic and the hierarchy inverted, because the MSW resonance happens in the $\bar\nu$ sector. This results in a  higher neutrino energy, as visible in figure  \ref{fi:figibdosc}, and in a larger number of events. This is clearly seen also in 
figure \ref{fi:repibd}, where we show the number of $\bar\nu_e$ interactions with protons that can be detected in LVD as a function of the $\bar\nu_e$ neutrino--sphere temperature.

\subsection{Charged current interactions with $^{12}C$}
\begin{figure}[b]
    \begin{center}
      \includegraphics[height=36pc]{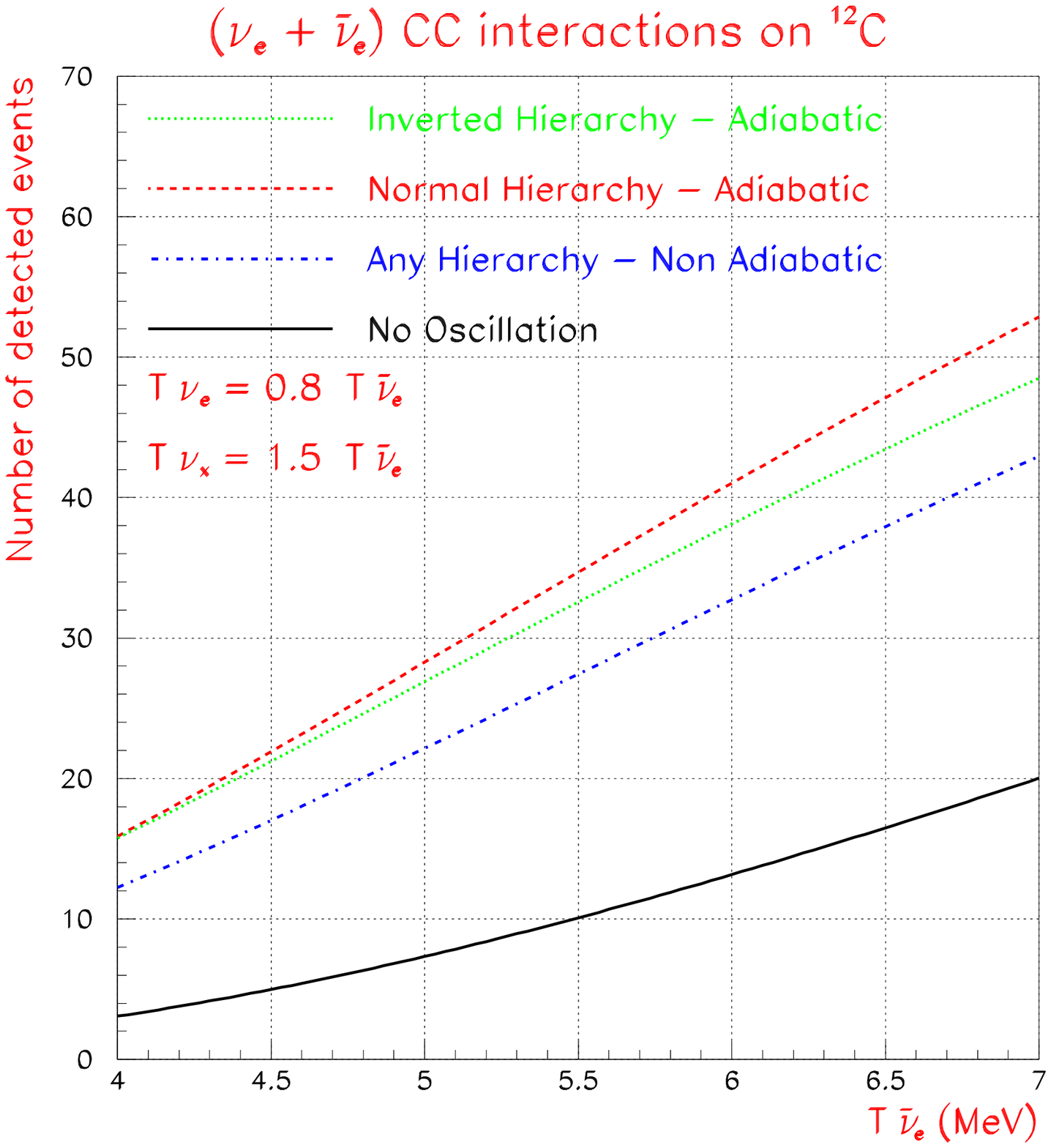}
    \end{center}
%    \vspace{-1.cm}
    \caption{Number of detectable $(\nu_e + \bar\nu_e)$ CC interactions with $^{12}$C expected in LVD as a function of the $\bar\nu_e$ neutrino--sphere temperature.}
    \label{fi:repcc}
%  \end{minipage}
\end{figure}

In figure \ref{fi:repcc} we show the expected number of  $(\nu_e + \bar\nu_e)$ charged current (CC) interactions with the $^{12}$C nuclei. The two contributions have the same signature in the detector if one looks for two high threshold signals in a time window of about $100~ms$, thus we consider them together. 
The conversion between the higher--energy non--electron neutrinos ($\nu_x$, $\bar\nu_x$ ) and the lower energy $\nu_e$, $\bar\nu_e$ , due to neutrino oscillation, increases the expected number of events. In the case of adiabatic transition the increase is even higher because at least one neutrino elicity state get a stronger contribution from the original $\nu_x$ (see eqq. \ref{eq:fenh}-\ref{eq:faeih}).

A strategy to statistically determine the separate amount of $\nu_e$ and $\bar\nu_e$ interactions, if a large number of CC interactions with $^{12}$C is detected, is described in \cite{AppC}. 

\subsection{Charged current interactions in the iron support structure} 
\begin{figure}[t] %checked 310106
%%  \begin{minipage}{.48\columnwidth}
    \begin{center}
      \includegraphics[height=36pc]{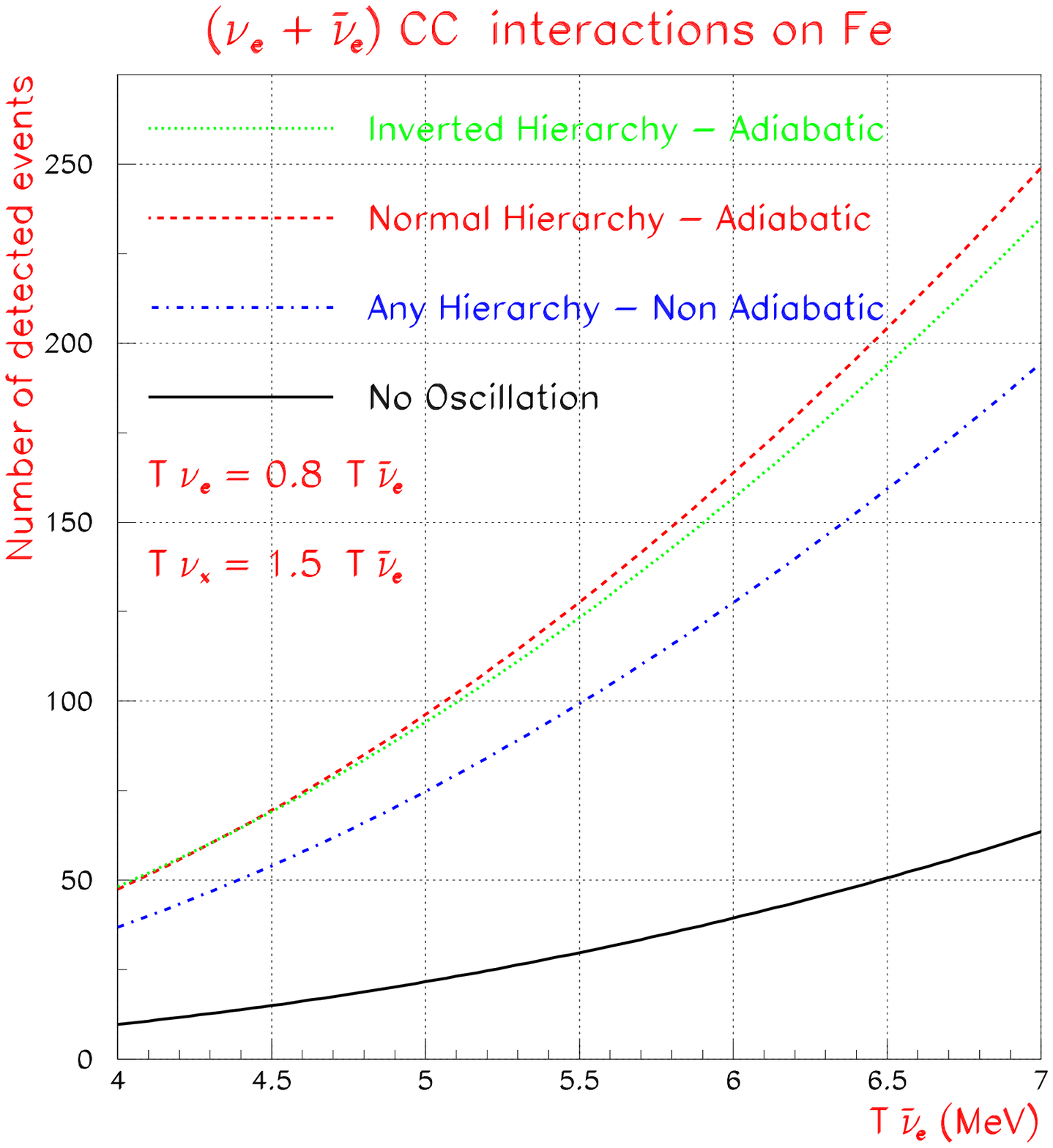}
    \end{center}
%    \vspace{-1.cm}
    \caption{Number of detectable $(\nu_e + \bar\nu_e)$ CC interactions with the iron of the support structure expected in LVD as a function of the $\bar\nu_e$ neutrino--sphere temperature.}
    \label{fi:repfe}
\end{figure}

\begin{figure}[t] %checked 010206
%%  \begin{minipage}{.48\columnwidth}
    \begin{center}
      \includegraphics[height=36pc]{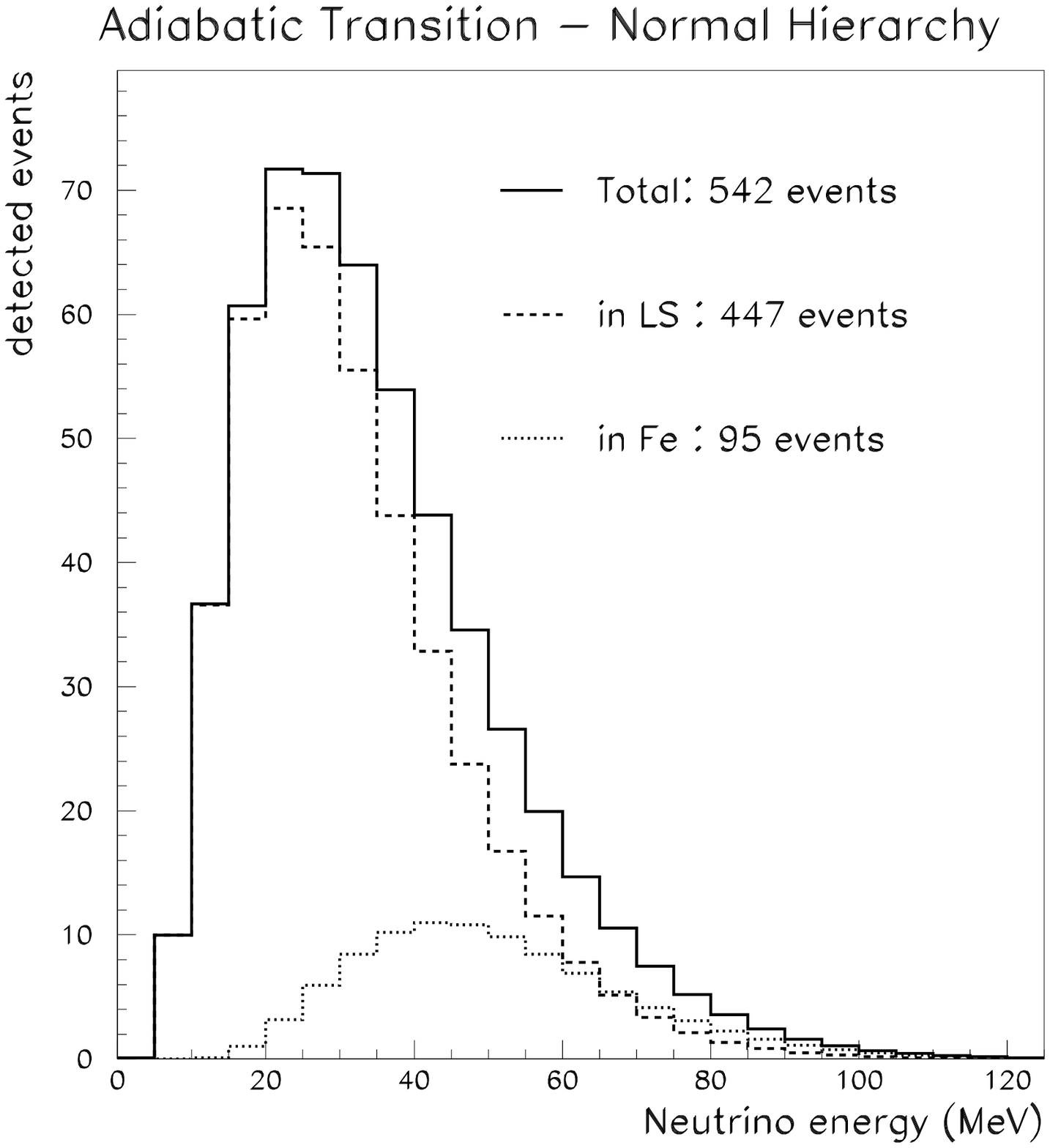}
    \end{center}
%    \vspace{-1.cm}
    \caption{Neutrino energy distribution of the events occurring in the liquid scintillator (dashed), in the iron support structure (dotted) and their sum (solid) in the LVD detector.}
    \label{fi:iron}
\end{figure}

An important contribution to the total number of events is also given by neutrino interactions in the iron support structure of the LVD detector. Given the rather high effective threshold (about $10$ MeV) and the increasing detection efficiency with the neutrino energy, they are concentrated in the high energy part of the spectrum ($E_\nu > 20$ MeV). Thus they are extremely sensitive to the neutrino energy spectrum and, indeed, to the oscillation parameters.\\
In figure \ref{fi:repfe} we show the dependence of the total number of detected $(\nu_e + \bar \nu_e)$ CC interactions with Fe to the $\anue$--sphere temperature 
, in the various oscillation scenarios.
In figure \ref{fi:iron} we show the contribution of $(\nu_e+\bar \nu_e)$ {\rm Fe} interactions on the total number of events. For the chosen supernova and oscillation parameters they are about $17\%$ of the total signal.
Indeed, they have to be considered in an accurate estimation of the expected events.

%  \end{minipage}
%  \hspace{1pc} %%%%% space between two figures

\subsection{Neutral current interactions on $^{12}C$}
\begin{figure}[b]
    \begin{center}
      \includegraphics[height=36pc]{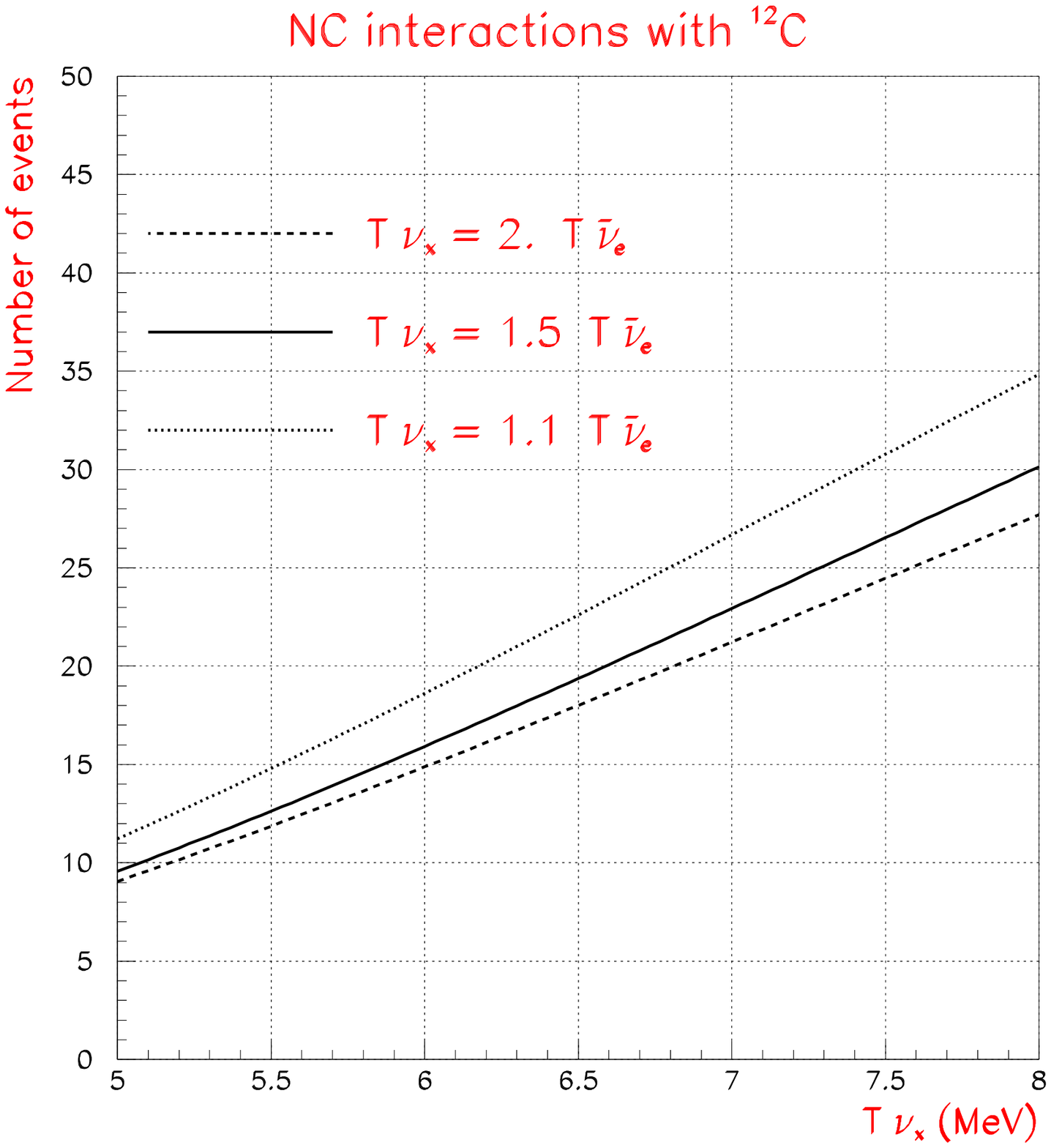}
    \end{center}
%    \vspace{-1.cm}
    \caption{Number of the detectable NC interactions of $\nu$ and $\bar\nu$ of all flavors with $^{12}$C expected in LVD as a function of the $\bar\nu_x$ neutrino--sphere temperature.} 
    \label{fi:kempnc}
%  \end{minipage}
\end{figure}

Neutral current interactions have the same cross section for all neutrino flavors, being thus insensitive to neutrino oscillations. 
Due to the high energy threshold ($15$ MeV) of the interaction, most of the detected events are given by the higher energy $\nux$. 
In principle, as shown in figure \ref{fi:kempnc} (solid line), NC with $^{12}$C could thus be used as a reference to identify the $\nux$--sphere temperature. 
However, the expected number of events depends 
also on the value of other astrophysical parameters (as it will be discussed in section \ref{se:ap}); for example, just changing the value of 
the other neutrino--sphere temperatures causes the variation in the number of events shown in figure \ref{fi:kempnc}, where two extreme values for the ratio $T_{\nu_x} / T_{\bar{\nu}_e}$  are chosen: $1.1$ (dotted) and $2.$ (dashed).

\subsection{Earth matter effect}
\begin{figure}[t]
    \begin{center}
      \includegraphics[height=36pc]{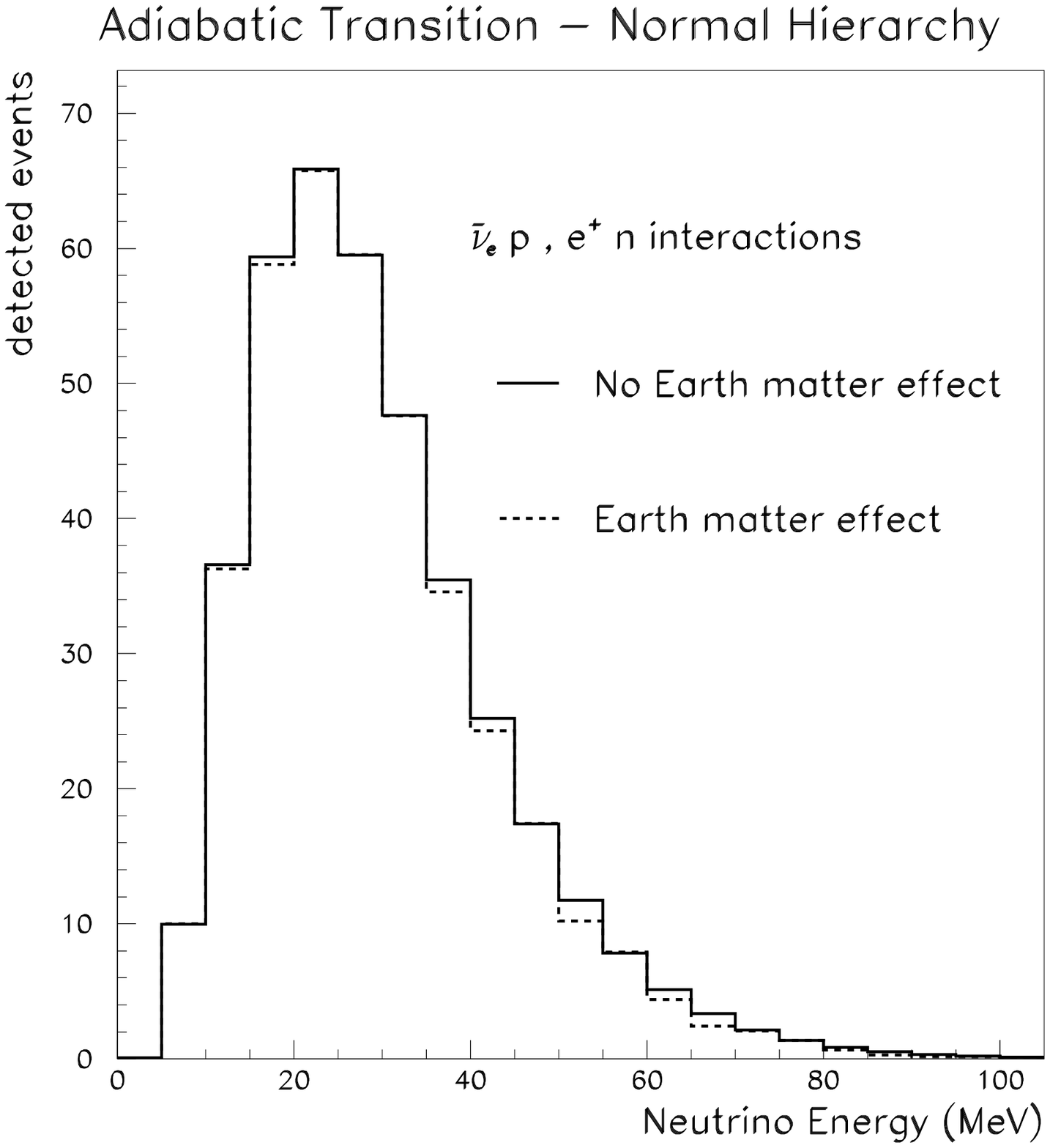}
    \end{center}
%    \vspace{-1.cm}
    \caption{Effect of the Earth matter in the $\bar\nu_e p,e^+ n$ signal in LVD, for neutrinos travelling through the Earth mantle (nadir angle $\theta_n=50^\circ$).}
    \label{fi:mswe}
\end{figure}

In order to measure the Earth matter effect at least two detectors in the world must detect the supernova neutrino signal, and one of them must be shielded by the Earth.
In figure \ref{fi:mswe} the effect of Earth matter in the inverse beta decay interaction channel, that is the most abundant and the cleanest one, is shown. The nadir angle is $\theta_n=50^\circ$, which corresponds to neutrinos travelling only through the mantle. The Earth matter effect produces a decrease in the number of detected neutrinos for particular neutrino energy, with a characteristic oscillating pattern. The effect is more relevant in the $\nu$ than in the $\bar \nu$ channel, so the effect in reaction {\em (1)} is quite weak (the weakness of the effect also depends on the rather high $\Delta m^2_{{\rm sol}} = 8 \cdot 10^{-5}~{\rm eV}^2$), but it could be detected if compared with a high statistic sample (i.e. with the Super-Kamiokande signal) or if a larger number of events is available (i.e. a closer supernova).

Even if the Earth matter effect is very difficult to be detected, we want to mark its importance. The comparison between the distorted and undistorted spectrum cancels many uncertainties coming from the astrophysical parameters that, as described in detail in the next section, introduce a large variability in the expected results. Then, the effect depends mainly on the $\theta_{12}$ and $\Delta m^2_{12}$ oscillation parameters, which are now known with good precision.
Moreover, even a null result carries many informations. For example, 
if the effect is not seen in the $\anue$ channel, it means that the transition is adiabatic and the hierarchy is inverted: in this case equation \ref{eq:faeih} becomes $F_{\bar e} = F_{\bar x}^0$, and it does not depend on the transition probabilities $P_{1e}$ and $P_{2e}$.
An exhaustive discussion about the implications of the Earth matter effect in the supernova neutrino signal can be found e.g. in \cite{CeciTucson}.

\section{Astrophysical parameter dependence}
\label{se:ap}
As discussed in section \ref{se:sn}, the emitted neutrino spectra depend on a large number of astrophysical parameters, whose values are not uniquely established by the actual MonteCarlo calculations. Thus it is very important to take into account the uncertainties in those parameters when estimating the total number of detected events.
 
A summary of the values of the astrophysical parameters used in our calculations is presented in table \ref{ta:param}, together with an estimation of their range of variability, as attempted e.g. in \cite{VissProbes}. The expected number of events in the various LVD detection channels and the mean energy of the detected $\anue~p$ events are shown in table \ref{ta:central}.

\begin{table}[H]
\begin{center}
\begin{tabular}{|l|c|c|c|}
\hline
Astrophysical parameter & Unit & Chosen value & Range of variability  \\
\hline
\hline
$D$: distance to the star & kpc          & $10$      & $0.2 \div 20$   \\
\hline
$E_b$: total energy emitted in $\nu$'s & $10^{53}$ erg & $3.$  & $2. \div 5.$ \\
\hline
$f_{\nue}$: fraction of $E_b$ taken by $\nue$ & & $1/6$ & $1/10 \div 1/4$ \\
\hline
$T_{\anue}$: $\anue$--sphere temperature  & MeV & $5.$ & $4. \div 7.$ \\
\hline
$T_{\nu_e} / T_{\bar{\nu}_e}$ & & $0.8$ & $ 0.5 \div 0.9 $ \\
\hline
$T_{\nu_x} / T_{\bar{\nu}_e}$ & & $1.5$ & $ 1.1 \div 2. $ \\
\hline
$\eta$: pinching parameter & & $0.$ & $0. \div 2.$ \\
\hline 
\end{tabular}
\end{center}
\vspace{0.5 cm}
 \caption{Astrophysical parameters values used in the calculations and their assumed uncertainties. \label{ta:param}}
\end{table}

\begin{table}[H] %checked
\begin{center}
\begin{tabular}{|l|c|c|c|c|}
\hline
  & No Oscillation & Non Adiabatic & Adiabatic NH & Adiabatic IH  \\
\hline
\hline
$\anue ~ p$   & 346. & \multicolumn{2}{c|}{391.} &   494. \\
\hline
$\langle E_{\anue} \rangle$ in $\anue ~ p$ & 25. MeV  & \multicolumn{2}{c|}{30. MeV}  & 37. MeV \\
\hline
 CC with $^{12}$C & 8. & 22. &   29. &    27. \\
\hline
 CC with $^{56}$Fe & 22. & 72. &   95. &    92. \\ 
\hline
 NC with $^{12}$C  & \multicolumn{4}{c|}{27}  \\
\hline
\end{tabular}
\end{center}
\vspace{0.5 cm}
 \caption{Expected results in the various LVD detection channels and in the mean energy of the detected $\anue~p$ events, calculated considering the chosen values of the astrophysical parameters, as given in table \ref{ta:param}.
 \label{ta:central}}
\end{table}

Our approach will be the following: we change the value of one parameter from the value listed in table \ref{ta:param} to the extreme lower and upper values, leaving the other astrophysical parameters unchanged and we show the fractional modification in the number of events and in the mean energy of the detected $\anue$ in $\anue ~p$ interactions.

The distance to the supernova ($D$) and the total energy emitted as neutrinos ($E_b$) appear in the time-integrated neutrino spectrum (equation \ref{eq:spectra}) as multiplier factors ($N_{ev} \propto E_b/D^2$). 
Thus, their uncertainty can greatly modify the expected number of events.
However, the use of the ratios of appropriate observables (i.e. inverse beta decay vs neutral current with carbon) allow to factorize them away. 
The energy spectra is not distorted by a change in $E_b$ nor $D$, so $<E_{\anue}>$ is unmodified.

The energy taken by each neutrino flavor is usually considered as equipartitioned; but differences up to a factor of two are allowed. If we consider $f_{\anue} = f_{\nue}$ and the normalization $( f_{\nue} + f_{\anue} + 4 f_{\nux}) =1$, we can choose $f_{\nue}$ as the only independent parameter. We show in table \ref{ta:frac} the fractional modifications corresponding respectively to the values $f_{\nue}=1/10 ~ (\rightarrow f_{\nux}=1/5)$ and $f_{\nue}=1/4  ~ (\rightarrow f_{\nux}=1/8)$.

\begin{table}[H] %checked
\begin{center}
\begin{tabular}{|l|c|c|c|c|}
\hline
  & No Oscillation & Non Adiabatic & Adiabatic NH & Adiabatic IH  \\
\hline
\hline
$\anue ~ p$     & $-40\%$ , $+50\%$ & \multicolumn{2}{c|}{$-18\%$ , $+22\%$} &  $+20\%$ , $-25\%$ \\
\hline
$\langle E_{\anue} \rangle$ in $\anue ~ p$ & $0\%$  & \multicolumn{2}{c|}{$+7\%$ , $-6\%$}  & $0\%$ \\
\hline
CC with $^{12}$C & $-40\%$ , $+51\%$  & $+8\%$ , $-10\%$  &  $+13\%$ , $-16\%$ & $+18\%$ , $-22\%$ \\
\hline
 CC with $^{56}$Fe & $-40\%$ , $+50\%$ & $+9\%$ , $-12\%$ &  $+14\%$ , $-17\%$ &  $+18\%$ , $-23\%$ \\ 
\hline
 NC with $^{12}$C  & \multicolumn{4}{c|}{$+13\%$ , $-16\%$}  \\
\hline
\end{tabular}
\end{center}
\vspace{0.5 cm}
 \caption{Fractional variations in the expected results if the energy fraction taken by $\nue$ is 1/10 (left value) or 1/4 (right value), with respect to the chosen value 1/6.
 \label{ta:frac}}
\end{table}

The values of the $\nue$, $\anue$ and $\nux$ neutrino--sphere temperatures
determine the energies of the incoming neutrinos. They are parametrized in terms of the $\anue$--sphere temperature $T_{\anue}$, and the ratios $T_{\nu_e} / T_{\bar{\nu}_e}$ and $T_{\nu_x} / T_{\bar{\nu}_e}$. The fractional variations when changing these three parameters are shown respectively in tables \ref{ta:tanue}, \ref{ta:rnuenuae}, \ref{ta:rnuxnuae}.

\begin{table}[H] %checked
\begin{center}
\begin{tabular}{|l|c|c|c|c|}
\hline
  & No Oscillation & Non Adiabatic & Adiabatic NH & Adiabatic IH  \\
\hline
\hline
$\anue ~ p$     & $-20\%$ , $+35\%$ & \multicolumn{2}{c|}{$-19\%$ , $+33\%$} &  $-17\%$ , $+29\%$ \\
\hline
$\langle E_{\anue} \rangle$ in $\anue ~ p$ & $-19\%$ , $+37\%$  & \multicolumn{2}{c|}{$-19\%$ , $+36\%$}  & $-19\%$ , $+37\%$ \\
\hline
CC with $^{12}$C & $-56\%$ , $+164\%$  & $-44\%$ , $+94\%$  &  $-43\%$ , $+85\%$ & $-41\%$ , $+79\%$ \\
\hline
 CC with $^{56}$Fe & $-55\%$ , $+193\%$ & $-51\%$ , $+164\%$ &  $-51\%$ , $+161\%$ &  $-49\%$ , $+152\%$ \\ 
\hline
 NC with $^{12}$C  & \multicolumn{4}{c|}{$-40\%$ , $+77\%$}  \\
\hline
\end{tabular}
\end{center}
\vspace{0.5 cm}
 \caption{Fractional variations in the expected results if the $\anue$ neutrino--sphere temperature is equal to $4$ MeV (left value) or $7$ MeV (right value), with respect to the chosen value $5$ MeV.
 \label{ta:tanue}}
\end{table}

\begin{table}[H]  %checked
\begin{center}
\begin{tabular}{|l|c|c|c|c|}
\hline
  & No Oscillation & Non Adiabatic & Adiabatic NH & Adiabatic IH  \\
\hline
\hline
$\anue ~ p$     &  \multicolumn{4}{c|}{$0\%$  } \\
\hline
$\langle E_{\anue} \rangle$ in $\anue ~ p$ &      \multicolumn{4}{c|}{$0\%$} \\    
\hline
CC with $^{12}$C & $-34\%$ , $+24\%$  & $-5\%$ , $+3\%$  &  $0\%$ , $0\%$ & $-4\%$ , $+2\%$ \\
\hline
 CC with $^{56}$Fe & $-31\%$ , $+21\%$ & $-3\%$ , $+2\%$ &  $0\%$ , $0\%$ &  $-3\%$ , $+2\%$ \\ 
\hline
 NC with $^{12}$C  & \multicolumn{4}{c|}{$-4\%$ , $+2\%$}  \\
\hline
\end{tabular}
\end{center}
\vspace{0.5 cm}
 \caption{Fractional variations in the expected results if the ratio between the $\nue$ and the $\anue$ neutrino--sphere temperature is equal to $0.5$ (left value) or $0.9$ (right value), with respect to the chosen value $0.8$.
 \label{ta:rnuenuae}}
\end{table}

\begin{table}[H]  %checked
\begin{center}
\begin{tabular}{|l|c|c|c|c|}
\hline
  & No Oscillation & Non Adiabatic & Adiabatic NH & Adiabatic IH  \\
\hline
\hline
$\anue ~ p$    & $0\%$ , $0\%$ & \multicolumn{2}{c|}{$-9\%$ , $+10\%$} &  $-24\%$ , $+25\%$ \\  
\hline
$\langle E_{\anue} \rangle$ in $\anue ~ p$ &   $0\%$ , $0\%$  & \multicolumn{2}{c|}{$-13\%$ , $+19\%$}  & $-25\%$ , $+31\%$ \\
\hline
CC with $^{12}$C & $0\%$ , $0\%$  & $-43\%$ , $+54\%$  &  $-48\%$ , $+60\%$ & $-50\%$ , $+59\%$ \\
\hline
 CC with $^{56}$Fe & $0\%$ , $0\%$ & $-52\%$ , $+106\%$ &  $-57\%$ , $+116\%$ &  $-59\%$ , $+116\%$ \\ 
\hline
 NC with $^{12}$C  & \multicolumn{4}{c|}{$-44\%$ , $+51\%$}  \\
\hline
\end{tabular}
\end{center}
\vspace{0.5 cm}
 \caption{Fractional variations in the expected results if the ratio between the $\nux$ and the $\anue$ neutrino--sphere temperature is equal to $1.1$ (left value) or $2.$ (right value), with respect to the chosen value $1.5$.
 \label{ta:rnuxnuae}}
\end{table}

%%%% pinching
The neutrino energy spectrum is presumably a black-body of the Fermi-Dirac type, but possible non-thermal effects are taken into account by introducing the parameter $\eta$ in equation \ref{eq:spectra}, the so-called ``pinching'' factor. A distribution with $\eta>0$ and the same average energy is, in fact, suppressed at low and high energies. In table \ref{ta:pinching} we show the fractional differences in the results when considering $\eta = 1$ or $\eta = 2$, in the simplified scenario where all the neutrino flavors are described by the same pinching factor. 

\begin{table}[H] %checked
\begin{center}
\begin{tabular}{|l|c|c|c|c|}
\hline
  & No Oscillation & Non Adiabatic & Adiabatic NH & Adiabatic IH  \\
\hline
\hline
$\anue ~ p$     & $+3\%$ , $+9\%$ & \multicolumn{2}{c|}{$+3\%$ , $+9\%$} &  $+3\%$ , $+8\%$ \\
\hline
$\langle E_{\anue} \rangle$ in $\anue ~ p$ & $+2\%$ , $+5\%$  & \multicolumn{2}{c|}{$+1\%$ , $+5\%$}  & $+2\%$ , $+5\%$ \\
\hline
CC with $^{12}$C & $+7\%$ , $+21\%$  & $+6\%$ , $+18\%$  &  $+6\%$ , $+18\%$ & $+6\%$ , $+17\%$ \\
\hline
 CC with $^{56}$Fe & $+7\%$ , $+20\%$ & $+6\%$ , $+19\%$ &  $+7\%$ , $+19\%$ &  $+6\%$ , $+18\%$ \\ 
\hline
 NC with $^{12}$C  & \multicolumn{4}{c|}{$+6\%$ , $+17\%$}  \\
\hline
\end{tabular}
\end{center}
\vspace{0.5 cm}
 \caption{Fractional variations in the expected results if the pinching parameter $\eta$ is equal to $1$ (left value) or $2$ (right value), with respect to the chosen value $\eta = 0$.
 \label{ta:pinching}}
\end{table}

We conclude that the sources of uncertainty in the astrophysical parameters which mostly affect the results are the partition of the available energy among the neutrino flavors and the values of the various neutrino--sphere temperatures. 
The largest variations in the expected signal (up to more than $100 \%$) are hence due to the poor (and hard to get) theoretical knowledge of the physics of the gravitational collapse, which will be hopefully improved at the occurence and detection of the next galactic supernova. 
With respect to the distance and the total released energy,
which appear as a $E_b/D^2$ multiplier factor,
 their
uncertainties certainly affect the signal, but,
even if not constrained by the observation, they can be factorized away by using appropriate observables (e.g. the ratio of the NC events and the $\anue ~p$ events).

\section{Summary and conclusions}
\label{se:sum}
The main aim of this paper was to show how neutrino oscillations 
affect the signal expected in the LVD detector at the occurence of the next galactic supernova.

The LVD detector has been described in its main components. 
It is able to detect neutrinos of all flavors, by studying them in the various CC and NC channels.
All the neutrino interactions that occur in the liquid scintillator as well as in the iron support structure have been 
studied in detail taking into account the neutrino energy threshold, cross section and detection efficiency.

We assumed a galactic supernova explosion at a typical
distance of $D = 10$~kpc, 
parametrized with a pure Fermi--Dirac energy spectrum ($\eta = 0$)
with a total energy $E_b = 3 \cdot
10^{53}$ erg and perfect energy equipartition $f_{\nu_e}=f_{\bar\nu_e}=f_{\nu_x}=1/6$; 
we fixed
$T_{\nu_x} / T_{\bar{\nu}_e} = 1.5$, 
$T_{\nu_e} / T_{\bar{\nu}_e}  = 0.8$
and $T_{\bar{\nu}_e} 
%\in \{4,~7\}~ 
=5~{\rm MeV}$.

We considered neutrino oscillations in the standard three-flavor scenario. The MSW effect has been studied in detail for neutrinos travelling through the supernova matter. We also considered the distortion in the expected neutrino spectra induced by a possible path inside the Earth before their detection.

For the chosen supernova parameters, it results that the expected number of events and their energy spectrum depend on the unknown oscillation parameters: the mass hierarchy and the value of $\theta_{13}$.

In particular, 
the inverse beta decay interactions ($\anue p, e^+ n$)
are highly sensitive to the mass hierarchy:
for adiabatic transition, the number of events
increases of $\sim 25 \%$ in the IH case, with respect to the NH one, since the detected $\anue$ completely come from the higher energy $\nux$. 
The mean energy of the detected positrons is correspondingly increased. 

The total number of $(\nu_e + \bar\nu_e)$ CC interaction with $^{12}$C nuclei is highly increased taking into account neutrino oscillations, because of their high energy threshold. For adiabatic transition the expected number of events is higher than the non adiabatic one, because at least one specie (between $\nu_e$ or $\bar\nu_e$) comes significantly from the original and higher--energy $\nux$ in the star. 
However, if it is not possible to discriminate between $\nu_e$ and $\bar\nu_e$, the normal and inverted hierarchy cases present similar results. Indeed, in the NH (IH) case, the increase in $\nue$ ($\anue$) is compensated by a decrease in $\anue$ ($\nue$).

The neutrino interactions with the iron of the support structure, which are studied in detail in this work, are also incread by the oscillations. The efficiency for the detection of the produced charged leptons and gammas in the active part of the detector has been calculated with a full simulation of the apparatus. The contribution of $(\nu_e+\bar \nu_e)$ {\rm Fe} interactions can be as high as $17\%$ of the total number of events (in the adiabatic NH case) and they contribute mostly to the high energy part of the spectrum.

With respect to the previous detection channels, the number of NC interactions with $^{12}$C nuclei does not depend on oscillations. In principle they could be used as a reference to identify the $\nux$--sphere temperature. However, this is partly limited by the uncertainties in the other astrophysical parameters.

We completed the calculations taking into account the effect of the passage of neutrinos through the Earth before their detection. This induces a characteristic modulation in the energy spectrum; however, given the expected number of events and the assumed oscillation parameters, the effect is quite weak.

In conclusion, for the choice of the astrophysical parameters adopted in this work, the expected signal of neutrinos in the LVD detector from a supernova core collapse greatly benefits of the neutrino oscillation mechanism, practically in all the possible detection channels, especially if the transition is adiabatic and the hierarchy inverted (since in LVD the most relevant signal is given by $\anue$).

However, being aware of the fact that the astrophysical parameters of the supernova mechanism are up to now not well defined, we performed the same calculations using different values of them. The resulting differences are in fact important; they are mainly due to the poor theoretical knowledge of the physics of the gravitational collapse.
This will be hopefully improved after the occurence and detection of the next galactic supernova, to which the LVD experiment can give a significant contribution, thanks to its cabability to observe and measure neutrino events of several types.

\appendix

\newpage
{\bf \Large Appendices}

\section{Oscillation probability in the Earth}
\label{AppA}
The neutrino flavor $|\nu_\alpha \rangle$ ($\alpha = e,\mu,\tau$) and mass $|\nu_i \rangle$ ($i=1,2,3$) eigenstates are related by: 
\be
|\nu_\alpha \rangle = U_{\alpha i}^* ~ |\nu_i \rangle 
\label{eq:vacmix}
\ee
(for antineutrino $U^*$ should be replaced by $U$)
where $U$ is the $3 \times 3$ mixing matrix in vacuum
\be
U =
\left(\begin{array}{ccc}
c_{12}\,c_{13}   & s_{12}\, c_{13}    & s_{13}\, e^{-i\delta} \\
-s_{12}\,c_{23}-c_{12}\,s_{23}\,s_{13}\,e^{i\delta} &
c_{12}\,c_{23}-s_{12}\,s_{23}\,s_{13}\,e^{i\delta} &
s_{23}\,c_{13}   \\
s_{12}\,s_{23}-c_{12}\,c_{23}\,s_{13}\,e^{i\delta} &
-c_{12}\,s_{23}-s_{12}\,c_{23}\,s_{13}\,e^{i\delta} &
c_{23}\,c_{13}
\end{array}
\right)\,.
\label{U3}
\ee
Since in the study of supernova neutrinos we are interested only in the $\nue$ and $\anue$ survival probabilities, the angle $\theta_{23}$ and the CP--violating phase $\delta$ do not matter.

The flavor eigenstates evolution in matter is governed by the following equation:

\begin{equation}
i \frac{d}{dt} ~ |\nu_\alpha \rangle = (\frac{1}{2E} U M^2 U^\dagger + V_{cc} )_{\alpha \beta}~ |\nu_\beta \rangle = H_{\alpha \beta} ~  |\nu_\beta \rangle
\end{equation}
where $M^2$ is the diagonal matrix of the squared neutrino masses and $V_{cc}$ is the matter induced potential $3 \times 3$ matrix 
\be
V_{cc} = \sqrt{2}\, G_F\, N_e
\left(\begin{array}{ccc}
1   & ~0   & ~~0 \\
0   & ~0   & ~~0 \\
0   & ~0   & ~~0 \\
\end{array}
\right)\,
\label{Vcc}
\ee
with 
$G_F$ the Fermi constant and $N_e$ the electron density in the crossed matter.

Diagonalizing $H$ we get 
\be
H = U_m D U_m^\dagger
\ee 
where $D$ is the diagonal matrix of the eigenvalues in matter and $U_m$ is the orthogonal matrix with the mass eigenstates in matter as column. 
We thus define $|\nu_i^m \rangle$, the neutrino eigenstates for the propagation in matter of constant density $N_e$, as 

\be
|\nu_i^m \rangle = (U_m^\dagger)_{i \alpha} ~ |\nu_\alpha \rangle 
\ee
and the evolution equation is 
\be
i \frac{d}{dt} ~ |\nu_i^m \rangle = D ~ |\nu_i^m \rangle 
\ee
The mass eigenstates at the time $t$ become
\be
|\nu_i^m (t) \rangle = \e^{-iDt} ~ |\nu_i^m (0) \rangle 
\ee
and the flavor eigenstates at the time $t$ are 
\be
|\nu_\alpha (t) \rangle =  
%U_m |\nu_i^m (t) \rangle = U_m  \e^{-iDt} ~ |\nu_i^m (0) \rangle =   
U_m  \e^{-iDt} U_m^\dagger~ |\nu_\alpha (0) \rangle =  S(t,0) ~ |\nu_\alpha (0) \rangle ~~
\label{eq:propag}
\ee

where $S(t,0) = U_m  \e^{-iDt} U_m^\dagger$ is the propagator of the flavor eigenstate $|\nu_\alpha \rangle$ from the time $t=0$ to the time $t$, in the matter of constant density $N_e$.

This result is valid if the neutrinos travel through a single layer of constant density described by $N_e$. Describing the density distribution as a series of steps, each of constant density, as it is for the Earth interior, we only need to replace the propagator $S(t,0)$ with the product of many propagators, one for each crossed density layer. Suppose, for example, that the neutrinos enter the layer $1$ with density $N_e^1$ at the time $t_1$ and exit at $t_2$ going through the layer $2$ with density $N_e^2$ until the time $t_3$ when they are detected.
In this case we can decompose the propagator as 
\be
S(t_3,t_1) = S_2(t_3,t_2) \cdot S_1(t_2,t_1) 
\ee
where each propagator has to be calculated as in eq. \ref{eq:propag} considering the particular density of the corresponding layer.

\begin{figure}[h]
    \begin{center}
      \includegraphics[height=24pc]{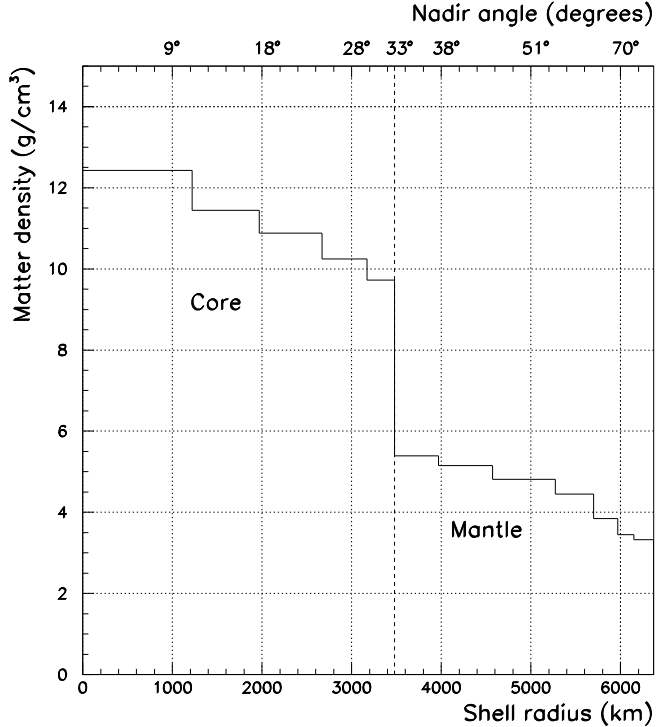}
    \end{center}
%    \vspace{-1.cm}
    \caption{Density of each layer inside the Earth in term of their radius (lower X axis). The nadir angle ($\theta_n$)corresponding to each shell radius is shown in the top level of the figure. The vertical dashed line at 3480 km ($\theta_n = 33^\circ$) represents the division between the Core and the Mantle inside the Earth, where a large change in density occurs.}
    \label{fi:prem}
%  \end{minipage}
\end{figure}

In our calculations the Earth interior has been divided into 12 layers of constant density, following the PREM model \cite{prem}; they are shown in figure \ref{fi:prem}. Two main regions can be defined: the Core (radius $<3500$ km, nadir angle $\theta_n < 33^\circ$) and the Mantle (outside). 

Now, since we are interested in calculating the probability for the mass eigenstate $j$, coming from the supernova, to be detected as flavor $\beta$, we get

\be
P_{(j \rightarrow \beta)} = | \langle \nu_\beta (0)  |\nu_j (t) \rangle |^2 = | \sum_{\alpha=e,\mu,\tau} U_{j \alpha}^\dagger ~ S(t,0)_{\beta \alpha} |^2
\ee

\begin{figure}[h]
\hspace{-2pc}
  \begin{minipage}{.4\columnwidth}
%    \begin{center}
      \includegraphics[height=20pc]{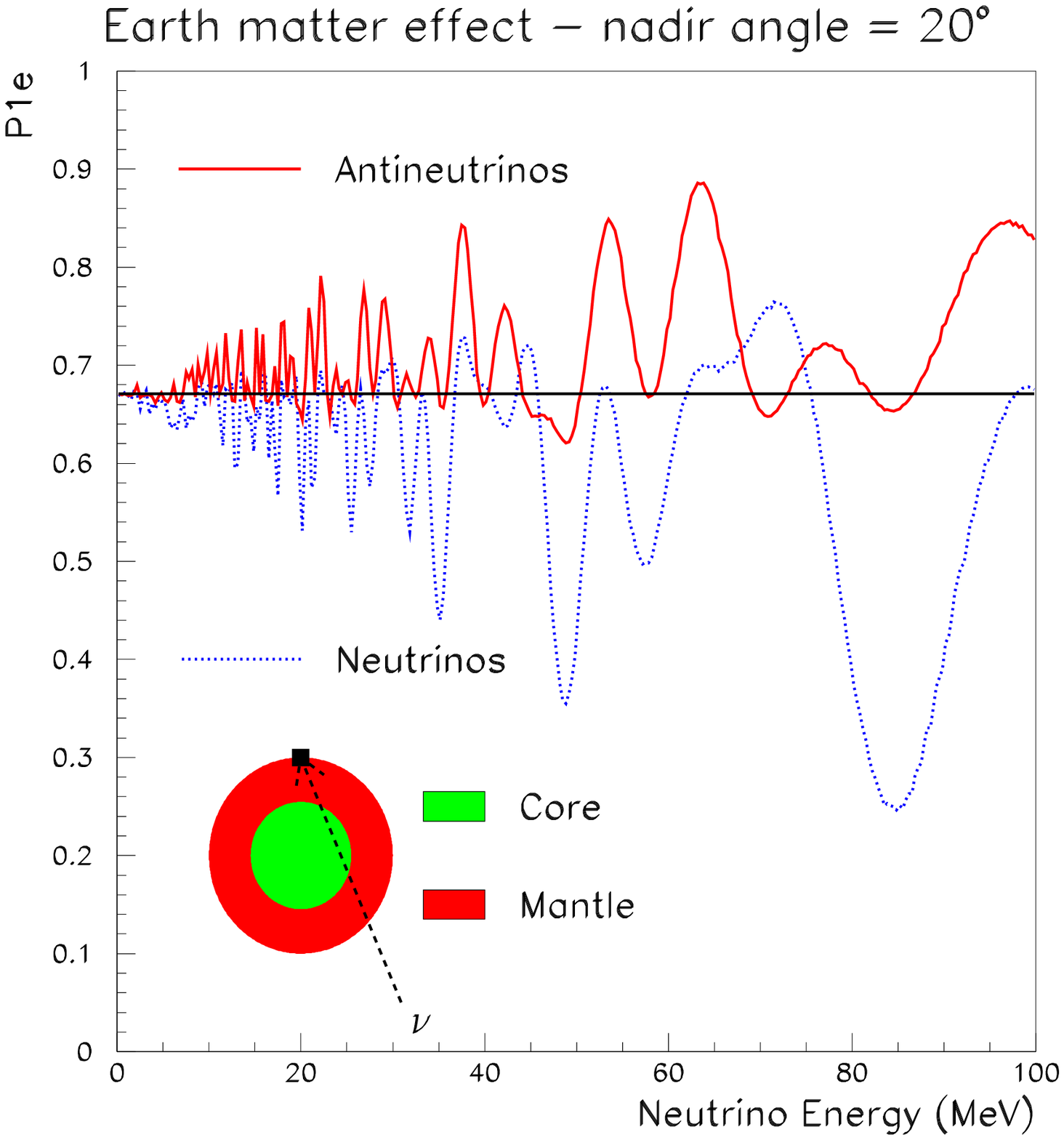}
%    \end{center}
%    \vspace{-1.cm}
%%\end{figure}
  \end{minipage}
  \hspace{3.5pc} %%%%% space between two figures
  \begin{minipage}{.4\columnwidth}
%%\begin{figure}[t]
%    \begin{center}
      \includegraphics[height=20pc]{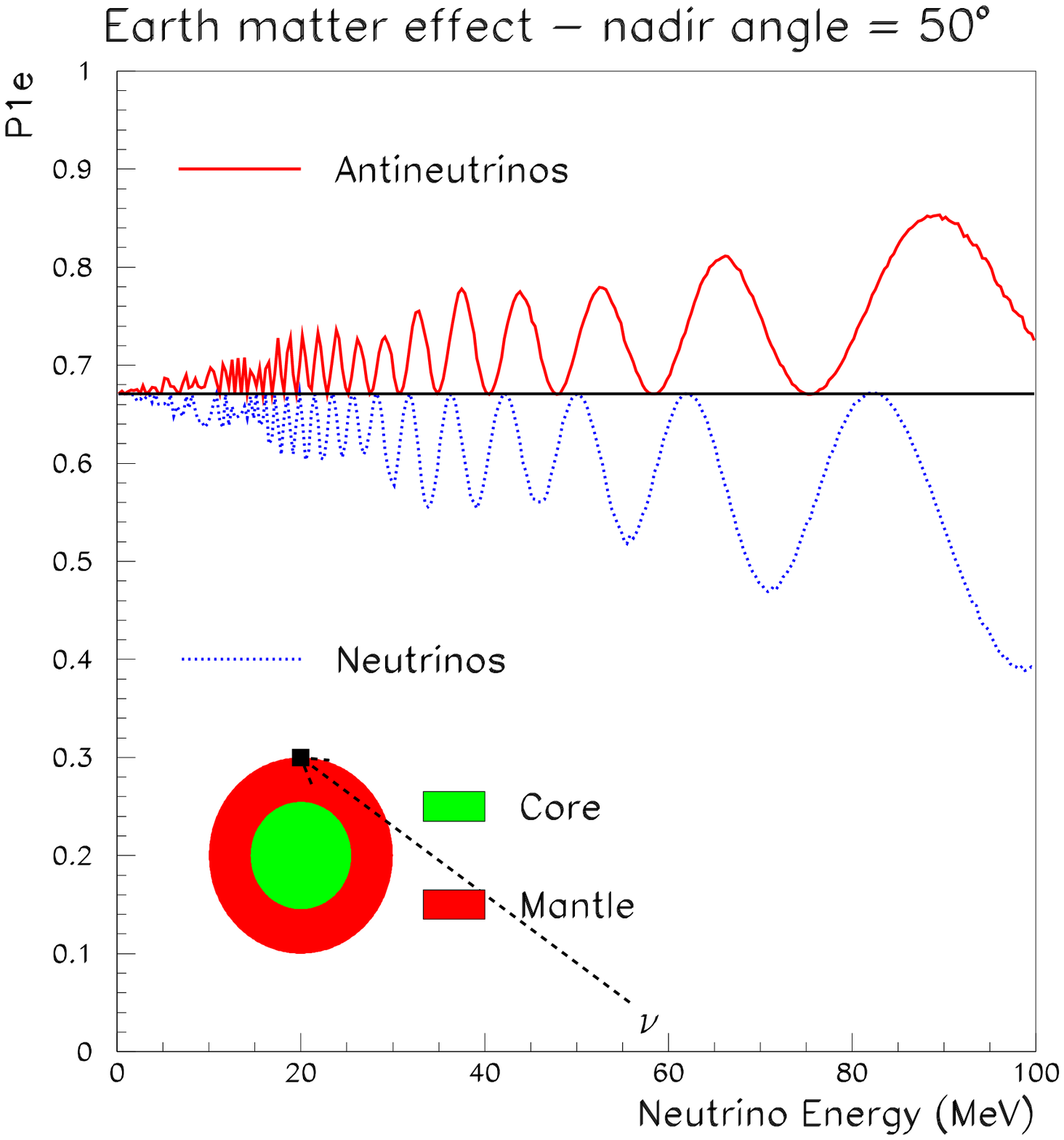}
%    \end{center}
%    \vspace{-1.cm}
  \end{minipage}
    \caption{$P_{1e}$ is the probability that the neutrino mass eigenstate $\nu_1$, coming from the supernova, is detected as $\nue$ after travelling through the Earth, as a function of the neutrino energy. The red solid line is for antineutrinos, while the blue dotted one is for neutrinos. The horizontal black solid line represents the value of $P_{1e}$ in the case of no Earth matter effect: $P_{1e}=\cos^2(\theta_{12})$. In the {\it left} plot the case of neutrinos travelling through the Mantle and the Core is shown ($\theta_n=20^\circ$), while in the {\it right} plot $\theta_n=50^\circ$ and the neutrinos go only through the Mantle, where the density is almost constant.}
    \label{fi:p1e}
\end{figure}

The results of the calculation of $P_{1e}$ are shown in figure \ref{fi:p1e}.
In the {\it left} plot the nadir angle is $20^\circ$ and the neutrinos go through both the Mantle and the Core. The interference between the two main density layers give rise to the complicated behavior of the probability $P_{1e}$. In the {\it right} plot $\theta_n = 50^\circ$ and the path is in the mantle only. The traversed density is thus almost constant and $P_{1e}$ becomes more regular.
An animated version of figure \ref{fi:p1e}, showing continuously what happen when the nadir angle changes, can be found in \cite{webSelvi}.

Similar plots can be obtained for the $P_{2e}$ probability. However, since $P_{3e}$ is very close to zero, for our purposes $P_{2e} = 1 - P_{1e}$ is a very good approximation.

When the neutrino path is only through the Mantle, the crossed density can be approximated as a single constant density step ($\rho \simeq 4 ~{\rm g/cm^3}$). In this particular case the $P_{1e}$ probabilities are exactly described by \cite{VissProbes}:

\be
P_{1e}=U^2_{e1}
-\frac{4\varepsilon \sin^2\theta_{12} \cos^2\theta_{12}}{
(1+\varepsilon)^2-4 \varepsilon\cos^2\theta_{12}
}\cdot
\sin^2\left( \frac{\Delta m^2_{12} L}{4 E} 
\sqrt{(1+\varepsilon)^2-4 \varepsilon\cos^2\theta_{12}}
\right)
\label{emeViss}
\ee

where 
$$
\varepsilon=
\frac{\sqrt{2} G_F N_e}{\Delta m^2_{12}/2 E}
\simeq 8\ \% \ 
\frac{\rho/(\mbox{4\ g/cm}^3)\cdot Y_e/(\mbox{0.5})
\cdot E/(20\mbox{ MeV})}{\Delta m^2_{12}/
(8\cdot 10^{-5}\mbox{ eV}^2)}
$$
For $\bar{\nu}_e$, just replace $\theta_{12}\to 90^\circ-\theta_{12}$.

\newpage
\section{Neutrino-iron interactions in LVD}
\label{AppB}
We consider the reaction $\nu_e\,^{56}\mathrm{Fe},^{56}\!\mathrm{Co}^*\ e^-$. It leads to excitation of analog $0^+$ and Gamow-Teller $1^+$ giant resonances (AR and GTR, respectively) in $^{56}\mathrm{Co}$ nucleus. The excitation of the AR is connected with Fermi transition. The ground state quantum numbers of $^{56}\mathrm{Co}$ are $4^+$, therefore the corresponding cross section with $^{56} \mathrm{Co}_{g.s.}$ in the final state is small, compared to AR and GTR excitation.

\begin{figure}[h]
    \begin{center}
      \includegraphics[height=24pc]{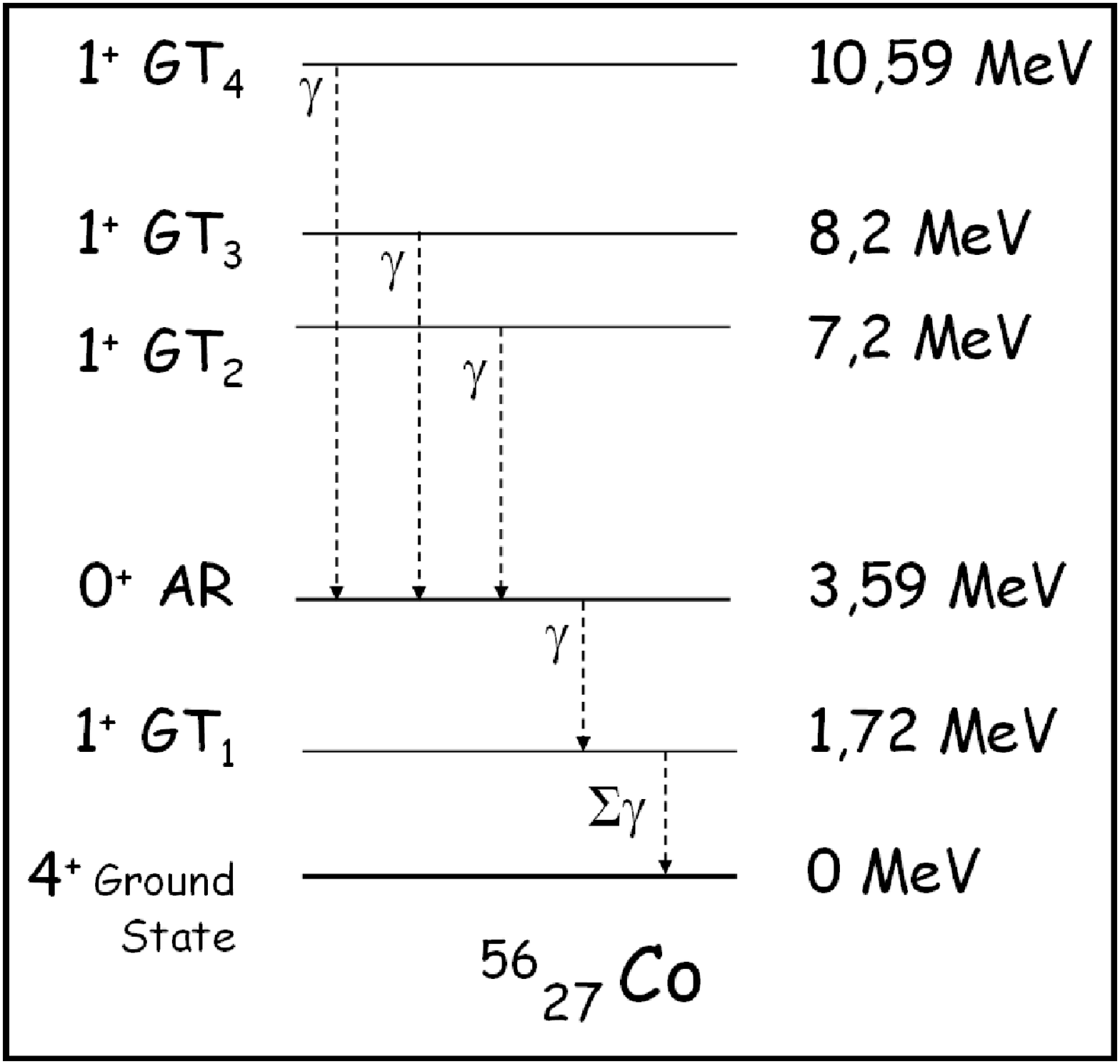}
    \end{center}
%    \vspace{-1.cm}
    \caption{Excitation levels of the 56 Cobalt nucleus.}
    \label{fi:Co}
%  \end{minipage}
\end{figure}

A simplified Cobalt nucleus level structure is shown in figure \ref{fi:Co}, where the $0^+$ AR is $3.59$ MeV over the ground state and the GTR take the other energy levels. Therefore, in our simulation, we take into account 5 possible channels \cite{Kurta}:

\begin{itemize}

\item AR) Cobalt is excited to the $0^+$ AR, at $3.59$ MeV. The products of the interaction are:
\begin{itemize}
\item the electron, with kinetic energy $E_{e^-} = E_{\nu_e} - (\Delta_{m_n} + E_{\rm level} + m_e)=E_{\nu_e}-8.156$ MeV, where $E_{\rm level}=3.59$ MeV and $\Delta_{m_n}=m_n^{\rm{Co}} - m_n^{\rm{Fe}} = 4.055~{\rm MeV}$,
\item one $1.87$ MeV gamma,
\item a gamma cascade, whose total energy is $1.72$ MeV\\
\end{itemize}

\item GT1) Cobalt is excited to the first Gamow-Teller resonance, at $1.72$ MeV. The products of the interaction are:
\begin{itemize}
\item the electron, with kinetic energy $E_{e^-} = E_{\nu_e}-6.286$ MeV,
\item a gamma cascade, whose total energy is $1.72$ MeV\\
\end{itemize}

\item GT2) Cobalt is excited to the second GTR, at $7.2$ MeV. The products of the interaction are:
\begin{itemize}
\item the electron, with kinetic energy $E_{e^-} = E_{\nu_e}-11.766$ MeV,
\item one $3.61$ MeV gamma,
\item one $1.87$ MeV gamma,
\item a gamma cascade, whose total energy is $1.72$ MeV\\
\end{itemize}

\item GT3) Cobalt is excited to the third GTR, at $8.2$ MeV. The products of the interaction are:
\begin{itemize}
\item the electron, with kinetic energy $E_{e^-} = E_{\nu_e}-12.766$ MeV,
\item one $4.61$ MeV gamma,
\item one $1.87$ MeV gamma,
\item a gamma cascade, whose total energy is $1.72$ MeV\\
\end{itemize}

\item GT4) Cobalt is excited to the fourth GTR state, at $10.6$ MeV. The products of the interaction are:
\begin{itemize}
\item the electron, with kinetic energy $E_{e^-} = E_{\nu_e}-15.166$ MeV,
\item one $7.$ MeV gamma,
\item one $1.87$ MeV gamma,
\item a gamma cascade, whose total energy is $1.72$ MeV
\end{itemize}

\end{itemize}

The neutrino-iron partial cross sections are shown in table \ref{ta:nuiron} for each considered excitation channel; they are taken from \cite{Kurta}.

\begin{table}[h]
 \caption{Neutrino-iron cross section for the various excitation level considered.}
 \label{ta:nuiron}
     \vspace{.3 cm}
\begin{center}
\begin{tabular}{|c|c|c|c|c|c|}
\hline
Neutrino energy (MeV) & \multicolumn{5}{c|}{ Cross section ($10^{-40}~{\rm cm}^2$) }     \\
 \hline
                      & GT1  &  AR  & GT2 & GT3 & GT4 \\
\hline
$10$         &$ 8.83~10^{-3}   $&$ 7.29~10^{-3} $& - & - & -  \\
$20$         &$ 9.58~10^{-2}   $&$ 1.96~10^{-1} $&$ 1.23~10^{-2}  $&$ 5.81~10^{-2}  $&$  6.76~10^{-2} $ \\
$30$         &$ 2.72~10^{-1}   $&$ 6.26~10^{-1} $&$ 5.49~10^{-2}  $&$ 2.96~10^{-1}  $&$  5.49~10^{-1} $ \\
$40$         &$ 5.36~10^{-1}   $&$ 1.29~10^{0}  $&$ 1.27~10^{-1}  $&$ 7.10~10^{-1}  $&$  1.48~10^{0}  $ \\
$50$         &$ 8.86~10^{-1}   $&$ 2.19~10^{0}  $&$ 2.28~10^{-1}  $&$ 1.30~10^{0}  $&$  2.84~10^0     $ \\
$60$         &$ 1.32~10^{0}    $&$ 3.32~10^{0}  $&$ 3.57~10^{-1}  $&$ 2.06~10^{0}  $&$  4.63~10^0     $ \\
$70$         &$ 1.84~10^{0}    $&$ 4.68~10^{0}  $&$ 5.15~10^{-1}  $&$ 2.99~10^{0}  $&$  6.85~10^0     $ \\
$80$         &$ 2.45~10^{0}    $&$ 6.26~10^{0}  $&$ 7.01~10^{-1}  $&$ 4.09~10^{0}  $&$  9.50~10^0     $ \\
$90$         &$ 3.14~10^{0}    $&$ 8.07~10^{0}  $&$ 9.15~10^{-1}  $&$ 5.36~10^{0}  $&$  1.26~10^{+1}  $ \\
$100$        &$ 3.90~10^{0}    $&$ 1.01~10^{1}  $&$ 1.16~10^{0}   $&$ 6.79~10^{0}  $&$  1.60~10^{+1}  $ \\
\hline
\end{tabular}
\end{center}
\end{table}

For each neutrino energy the excitation channel is sampled accordingly to its relative weight to the total cross section. The reaction products are generated uniformly in the LVD iron support structure. The electron and the gammas are simulated starting from the same generation point and their directions are chosen uniformly in the whole solid angle, without any correlation between them.

We used a GEANT3 simulation of the LVD detector, where the liquid scintillator and the iron support structure are described in detail. The particles of the electromagnetic showers are tracked through the various materials until their energy is smaller than $100$ keV, the lowest possible value allowed by the simulation program.

\begin{figure}[h]
%%  \begin{minipage}{.48\columnwidth}
    \begin{center}
      \includegraphics[height=30pc]{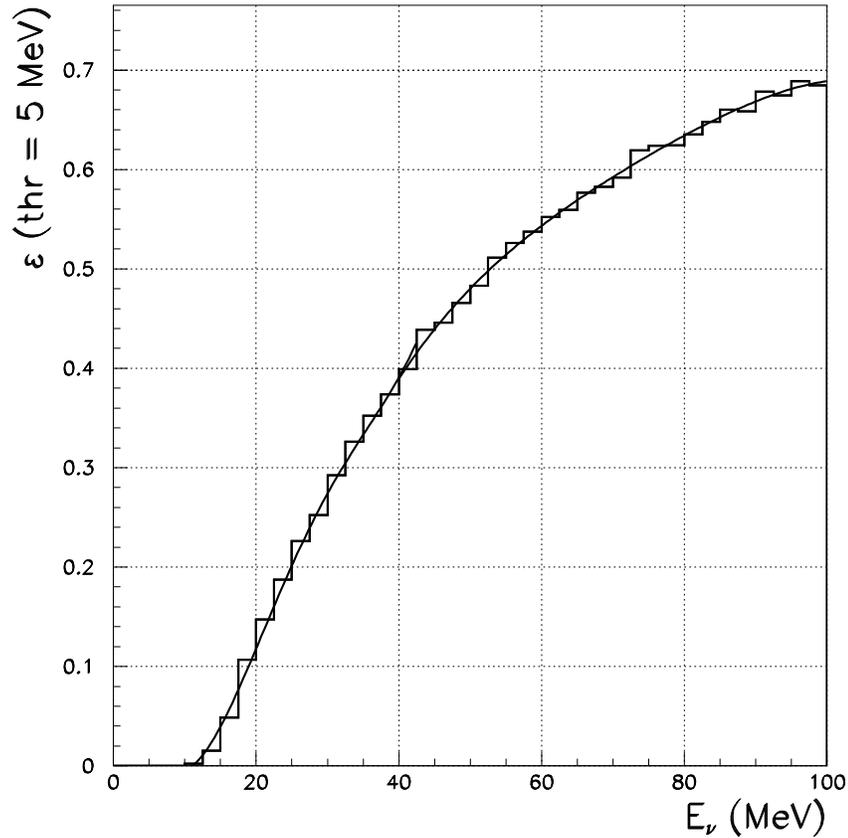}
    \end{center}
%    \vspace{-1.cm}
    \caption{Detection efficiency of neutrino-iron interaction for an energy threshold of the scintillation counters of $5$ MeV.}
    \label{fi:effFe}
\end{figure}

We define the efficiency as the ratio of the number of events where at least one scintillation counter detects a signal over its threshold and the total number of generated events.
The energy resolution of the detector is taken into account in the simulation.
Setting the energy threshold of all the scintillation counters to $5$ MeV, the resulting efficiency is shown in figure \ref{fi:effFe}.

\begin{figure}[h]
%%  \begin{minipage}{.48\columnwidth}
    \begin{center}
      \includegraphics[height=30pc]{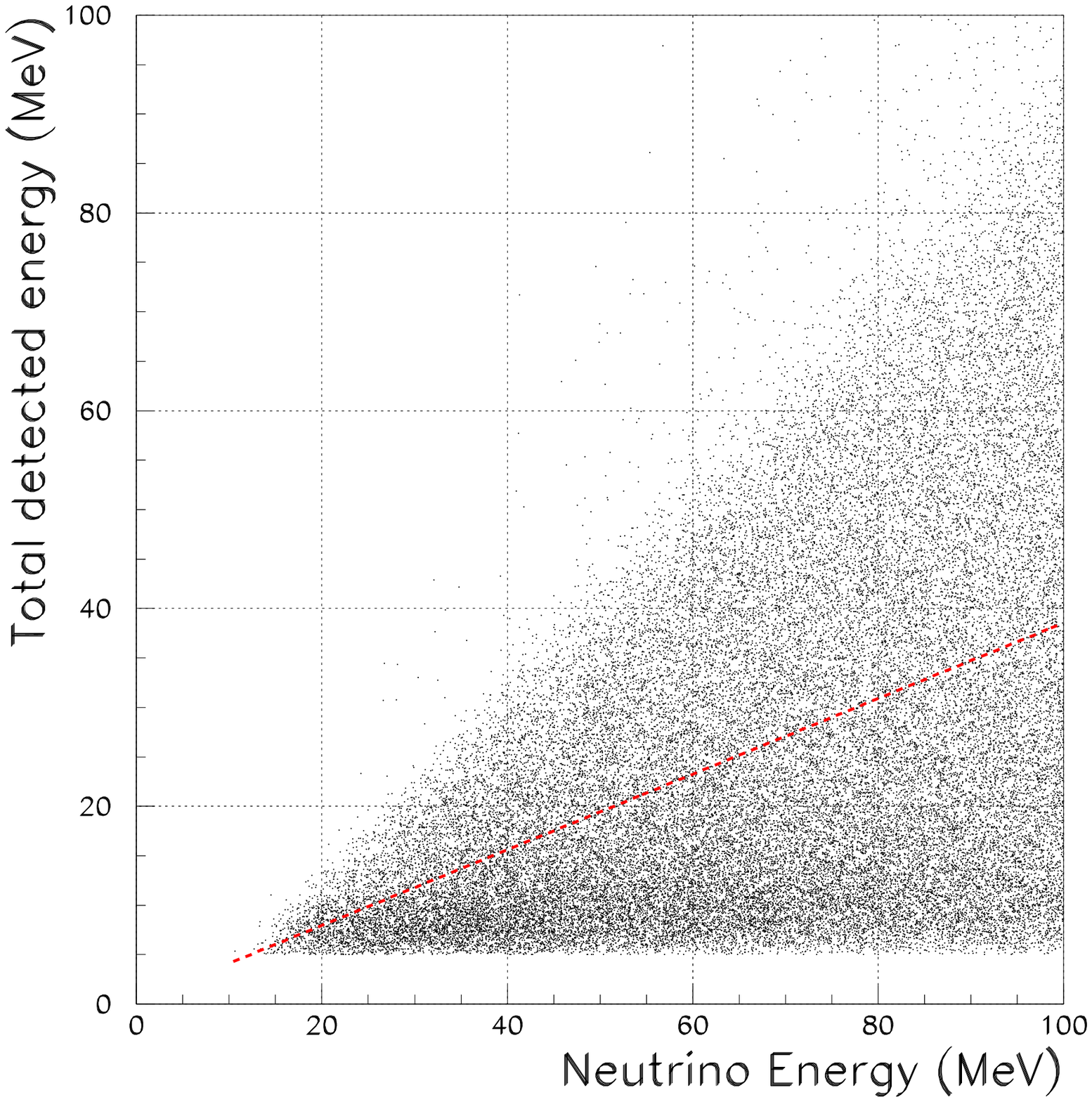}
    \end{center}
%    \vspace{-1.cm}
    \caption{Scatter plot of the total detected energy with respect to the incoming neutrino energy. The superimposed line is the relation between the average detected energy and the neutrino energy.}
    \label{fi:ecor}
\end{figure}

The total energy detected in the liquid scintillator is very weakly correlated to the neutrino energy, as shown in the scatter plot of figure \ref{fi:ecor}.
On average, the total energy detectable is $E_d \simeq 0.4 \times E_\nu$, but the spread over the mean value is very large.

\end{document}